\documentclass[superscriptaddress,aps,pra,twocolumn,showpacs,floatfix]{revtex4-2}
\usepackage[utf8]{inputenc}
\usepackage{graphicx,amsmath,amsfonts,amssymb}
\usepackage{color}
\usepackage[colorlinks=true, allcolors={blue}]{hyperref}
\usepackage{graphicx}
\usepackage{epstopdf}
\usepackage{float}
\usepackage{placeins}

\newcommand{\orcid}[1]{\href{https://orcid.org/#1}{\includegraphics[width=7pt]{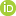}}}

\usepackage{mathtools, amssymb, mathrsfs, dsfont}

\usepackage{fancyhdr}
\usepackage{lipsum}

\usepackage[labeled]{multibib}
\newcites{S}{References}

\begin{document}

\preprint{APS/123-QED}

\title{Title}

\title{Quantum-State Texture Dynamics: Theory and Experiment}

\author{Carlos H. S. Vieira\,\orcid{0000-0001-7809-6215}}
\email[Corresponing author: ]{carloshsv09@gmail.com}
\affiliation{Centro de Ci\^{e}ncias Naturais e Humanas, Universidade Federal do ABC,
Avenida dos Estados 5001, 09210-580 Santo Andr\'e, S\~{a}o Paulo, Brazil}
\affiliation{Department of Physics, State Key Laboratory of Quantum Functional Materials,
and Guangdong Basic Research Center of Excellence for Quantum Science,
Southern University of Science and Technology, Shenzhen 518055, China}

\author{Xinfang Nie}
\email[Corresponding author: ]{niexinfang@quantumsc.cn}
\affiliation{Quantum Science Center of Guangdong-HongKong-Macao Greater Bay Area, Shenzhen 518045, China}
\affiliation{Department of Physics, State Key Laboratory of Quantum Functional Materials,
and Guangdong Basic Research Center of Excellence for Quantum Science,
Southern University of Science and Technology, Shenzhen 518055, China}

\author{Dawei Lu}
\email[Corresponding author: ]
{ludw@sustech.edu.cn}
\affiliation{Department of Physics, State Key Laboratory of Quantum Functional Materials,
and Guangdong Basic Research Center of Excellence for Quantum Science,
Southern University of Science and Technology, Shenzhen 518055, China}
\affiliation{Quantum Science Center of Guangdong-HongKong-Macao Greater Bay Area, Shenzhen 518045, China}

\author{Fernando Parisio\,\orcid{0000-0001-5818-8366}}
\email[Corresponding author: ]{fernando.parisio@ufpe.br}
\affiliation{Departamento de F\'{\i}sica, Centro de Ci\^encias Exatas e da Natureza, Universidade Federal de Pernambuco, Recife, Pernambuco
50670-901, Brazil}

\begin{abstract}
Quantum-state texture (QST) has found applications in several fields from quantum foundations and computation to quantum criticality. However, a general theory of QST dynamics under arbitrary physical processes remains unavailable, limiting both its practical application and experimental exploration. Here, we demonstrate that the QST response to an arbitrary finite-dimensional channel is fully encoded in the dual evolution of a single reference state. This description yields necessary and sufficient conditions for texture preservation and implies exact conservation under all free-unital dynamics. Using a nuclear magnetic resonance quantum processor, we experimentally verify these predictions across distinct channel classes. Furthermore, we show that local QST measurements provide an operational signature of entangling gates in circuit layers. Our results establish quantum-state texture as a resource and a practical diagnostic tool in quantum information processing.
\end{abstract}

\maketitle


\section{Introduction}\label{sec:intro}

When a physical quantity becomes relevant for performing useful tasks, it is viewed as a ``resource'' in a broad sense. Formalizing this notion in the non-classical realm gives rise to quantum resource theories, which provide a framework for quantifying, protecting, and exploiting quantum phenomena in information science \cite{gour}. 
These theories have helped characterize essential capabilities such as entanglement \cite{RevModPhys.81.865}, non-stabilizerness \cite{Veitch_2014}, and, of particular interest for the present work, quantum coherence \cite{l1}. 

Coherence is the fundamental property behind most non-classical phenomena, and, as such, its bare quantification may not be sufficient to fully explore its foundational and technological implications. Relevant complementary characterizations include genuine multi-level coherence \cite{MultiCoh} and imaginarity \cite{imag1,imag}.
Within this context, quantum-state texture (QST) has emerged as a basic resource \cite{texture} with a direct geometric interpretation: it measures the dissimilarity between an arbitrary density matrix and a flat reference state for a given basis. This quantity has proven versatile in various scenarios. For example, it serves as a sensitive probe for characterizing quantum phase transitions and criticality \cite{aditi,lucas} and helps to identify unknown gates \cite{texture, fixed-point} in quantum circuits. Beyond these uses, recent work has introduced generalizations of the original concept \cite{fixed-point,block} as well as new QST measures \cite{measure1,measure2,measure3,measure4,measure5,measure6} and established connections between QST and other key resources, including purity, entanglement, and non-stabilizerness \cite{aditi}. Furthermore, QST has been linked to quantum battery capacity \cite{bateries}, relativistic quantum field theory \cite{relativity,relativity2,relativity3,relativity4}, the principle of superposition \cite{sup}, and general frameworks for resource quantification \cite{fec}.

Despite these rapid theoretical developments, a fundamental gap remains: How QST transforms under general physical operations, including open-system quantum channels. Because a comprehensive dynamical theory of QST is lacking, engineering, protecting, or using this resource in general practical scenarios remains a challenge to be addressed on a case-by-case basis. Experimentally, the situation is equally unexplored, with no studies addressing QST so far.

In this work, we address both issues by establishing a dynamical theory of QST under quantum channels and by reporting the first experimental investigation on QST. Using nuclear-spin qubits and nuclear magnetic resonance (NMR) techniques~\cite{Ivan_book,JONES202449, Mahesh_review,Lu2016,Vieira2023}, we probe QST transformations across representative quantum channels. We further introduce and demonstrate a layer-resolved protocol in which local QST measurements directly signal entangling interactions, bypassing full quantum process tomography (QPT). Our findings establish QST as a tangible and operationally accessible resource.

\section{QST under general channels}
Given a fixed basis, hereafter denoted by $\{ |i \rangle\}$, 
the essentials of the QST resource theory are as follows \cite{texture}. A single pure state corresponds to the resourceless set $f_1\equiv |f_1\rangle \langle f_1| \in {\cal B(H)}$, with $ |f_1\rangle \equiv \frac{1}{\sqrt{D}}\sum_{i=1}^D|i\rangle$, where $D$ is the dimension (assumed to be finite) of the Hilbert space ${\cal H}$. We refer to this state as flat or textureless. 
The related free operations are completely positive and trace-preserving (CPTP) maps $\Lambda$, effected by Kraus operators, $\Lambda(\varrho)=\sum_nE_n\varrho E^{\dagger}_n$, for which $\sum_n E_n^{\dagger}E_n=\mathds{1}$. These free maps cannot create texture, and thus $f_1$ must be a fixed point: $\Lambda(f_1)=f_1$. Given an arbitrary texture measure, ${\cal T}$, it must be such that  {\bf (i)} ${\cal T}(f_1)=0$ and  {\bf (ii)} ${\cal T}(\varrho)\ge {\cal T}(\Lambda(\varrho))$ for all $\varrho \in {\cal B(H)}$.
Finally, we assume {\bf (iii)} convexity 
: ${\cal T}(\sum p_i \varrho_i) \le \sum p_i {\cal T}(\varrho_i)$. 

The grand sum of  $\Sigma(\varrho)$ is defined as
\begin{equation}
\Sigma(\varrho)\equiv D {\rm Tr}(\varrho f_1) = D \langle f_1|\varrho|f_1 \rangle= \sum_{i, j=1}^D\varrho_{ij}.
\label{eq01}
\end{equation}
For an arbitrary state $\varrho$, the rugosity
$\mathfrak{R}(\varrho)=-\ln \left(\Sigma(\varrho)/D\right) \in [0, \infty)$, has been shown to satisfy {\bf(i)}-{\bf (iii)}, being also additive under tensoring.
All results in this work are valid for any texture measure that depends solely on $\Sigma$, such as rugosity. Importantly, since for any state the contribution of the diagonal entries is fixed by ${\rm Tr}(\varrho)=1$, $\Sigma(\varrho)$ provides relevant information about the coherences, and so does any monotone based on $\Sigma$. 
We remark that $\Sigma(\varrho)$ is a directly measurable quantity, because $\Sigma(\varrho)/D={\rm Tr}(\varrho |f_1\rangle \langle f_1|)=P_{f_1}$ is the probability of obtaining $|f_1\rangle$ in a measurement of any observable that has $|f_1\rangle$ as one of its outcomes. This greatly facilitates the experimental determination of $\Sigma(\varrho)$, avoiding the need for full quantum-state tomography, as the last equality in Eq. (\ref{eq01}) would suggest.

Our purpose is to investigate the action of general CPTP maps $\Gamma$ on the QST of an arbitrary state $\varrho$, that is, how the grand sum of the transformed state, $\Sigma(\Gamma(\varrho))$, can be expressed.
Let us start with our most general result.

{\bf Proposition 1}: Let $\varrho$ be an arbitrary state in a finite-dimensional Hilbert-Schmidt space and $\Gamma$ a general CPTP map effected by the Kraus operators $\{K_j\}$. Then, the grand sum of $\Gamma(\varrho)=\sum_jK_j \varrho K_j^{\dagger}$ is given by:
\begin{equation}
\label{general}
\Sigma(\Gamma(\varrho)) = D {\rm Tr}(\tilde{\Gamma}_1 \varrho),
\end{equation}
where $D$ is the dimension of ${\cal H}$, $\tilde{\Gamma}$ is the dual map to $\Gamma$, and $\tilde{\Gamma}(f_1)\equiv \tilde{\Gamma}_1= \sum_jK^{\dagger}_j f_1 K_j$. The previous result can also be expressed as $\Sigma(\Gamma(\varrho))= D \Sigma(\tilde{\Gamma}_1^{\rm T} \odot \varrho)$, where T denotes transposition and $\odot$ the Hadamard product [given two matrices of the same dimension $A$ and $B$, then $(A \odot B)_{ij}\equiv A_{ij}B_{ij}$].  The demonstrations of all propositions in this work are given in Supplemental Material (SM)~\cite{supp_material}. 
Importantly, the QST resource theory fixes $f_1$, which is independent of the input state. Thus, for a given channel $\Gamma$, the complete QST response across arbitrary inputs is encoded in the single Hermitian operator $\tilde{\Gamma}(f_1)$, reducing the required characterization from $O(D^{4})$ parameters for a generic quantum channel to $O(D^{2})$.

Now we focus on a relevant subset of free operations: Those that protect QST. Although free operations in arbitrary resource theories can generally deplete resources, we show that, for a large class of channels, free operations strictly preserve QST.
By definition, free operations are CPTP maps for which $f_1$ is a fixed point, $\Lambda(f_1)= \sum_jE_j f_1 E_j^{\dagger}=f_1$, which, since $f_1$ is a pure state, implies $E_j f_1 E_j^{\dagger}=a_jf_1$, $\sum_j a_j=1$. 

{\bf Proposition 2}: An arbitrary channel $\Lambda$ is QST-preserving if and only if $\tilde{\Lambda}_1=f_1$, that is, if $f_1$ is also a fixed point of the dual map $\tilde{\Lambda}$. Also, for any QST-preserving map, all Kraus operators commute with $f_1$.

By a QST-preserving map, we mean a channel that leaves the texture of any input state invariant. Therefore, every QST-preserving map has $f_1$ as a common fixed point of both the channel and its dual, $\Lambda(f_1)=\tilde{\Lambda}(f_1)=f_1$.

{\bf Proposition 3}: Any free unital CPTP map necessarily preserves quantum-state texture.

 Interestingly, we show in SM that for a two-dimensional Hilbert space (qubit), a free CPTP map preserves QST if and only if it is unital. Note also that there are non-unital maps which preserve QST. In addition, unital channels may decrease QST for some states, but the previous result indicates that such maps cannot be free operations. Put differently, if a unital channel is such that $\Sigma(\Gamma(\varrho))>\Sigma(\varrho)$ for some $\varrho$, then there must exist a state $\sigma$ such that $\Sigma(\Gamma(\sigma))<\Sigma(\sigma)$. In fact, this observation has a rigorous notion of balancedness behind it.

\begin{figure}
\centering
\includegraphics[width=0.8\columnwidth]{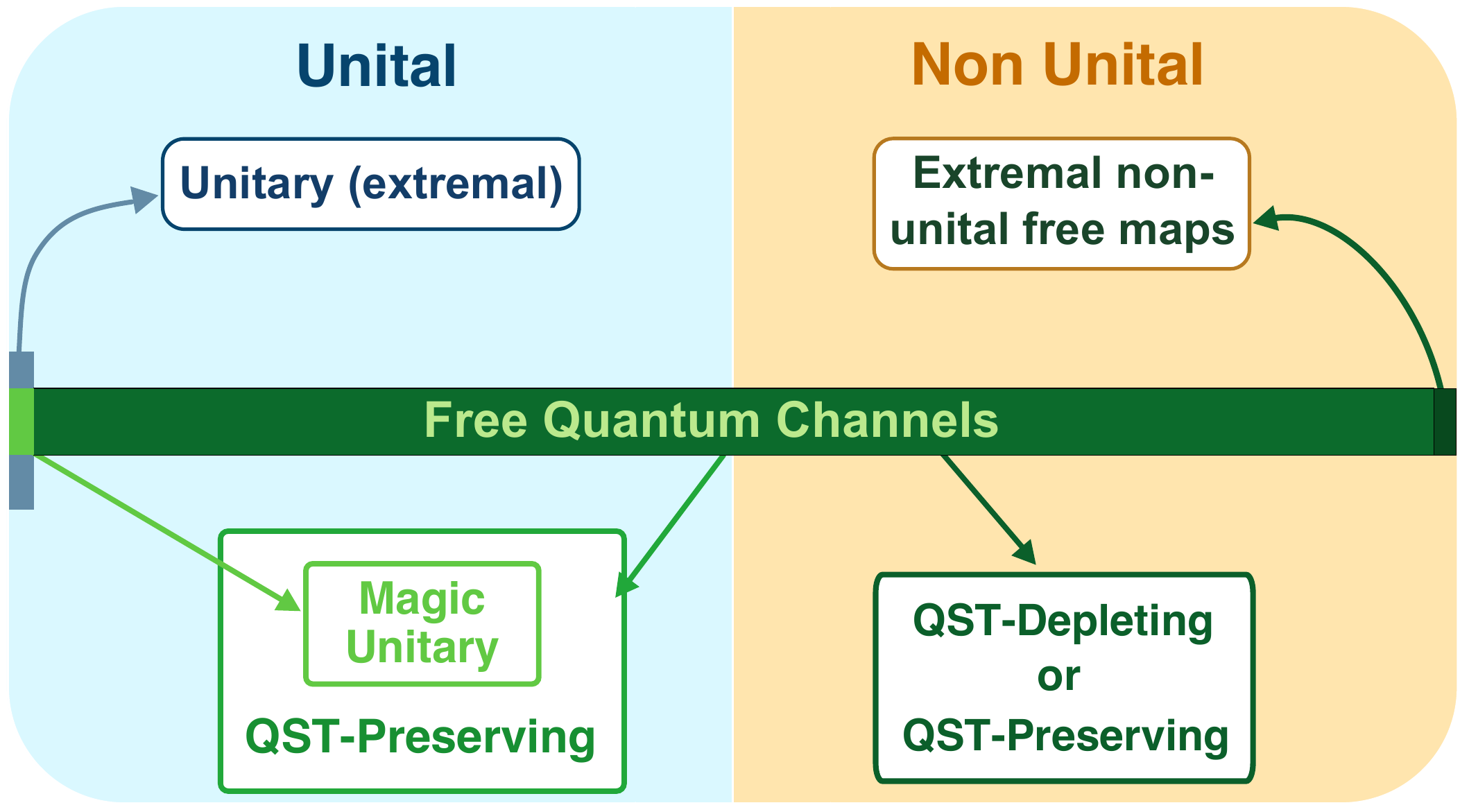}
\caption{Schematic classification of quantum channels from the perspective of QST theory. In the unital sector, unitary channels are extremal maps. 
The set of free channels in this sector is QST-preserving and contains magic unitary operations. In the non-unital sector, free maps may deplete QST.}
\label{fig1}
\end{figure}

{\bf Proposition 4} (Haar balancedness): An arbitrary unital map,  $\Gamma$, preserves the average grand sum with respect to the Haar measure. That is,

\begin{equation}
\label{HB}
\overline{\Sigma}_{\rm in}\equiv \int d\Omega  \Sigma({\psi_{\rm in}})  = \int d\Omega \Sigma(\Gamma({\psi_{\rm in}}))\equiv\overline{\Sigma}_{\rm out}=1,
\end{equation}

where $ d\Omega$ denotes the Haar integration measure. 

In contrast, non-unital channels can easily violate the balancedness condition. The previous proposition considerably generalizes a result presented in the supplemental material of \cite{texture} (corresponding to $\overline{\Sigma}_{\rm out}=1$ for unitary operations over a qubit). We have used this statement to distinguish entangling from non-entangling unitaries in unknown quantum circuit layers. Here, too, the average grand sum can serve this purpose in the expanded scenario of qudits under unital operations. However, sampling over all pure states may be a practical bottleneck. In what follows, we prove a much simpler statement, which leads to an experimentally accessible protocol to distinguish entangling from non-entangling unital maps.

{\bf Proposition 5} (Basis balancedness):  An arbitrary unital map,  $\Gamma$, preserves the average grand sum over any orthonormal basis. That is,

\begin{equation}
\label{BB}
\langle \Sigma_{\rm in} \rangle \equiv\sum_{j=1}^D\frac{\Sigma(u_j)}{D}=
\sum_{j=1}^D\frac{\Sigma(\Gamma(u_j))}{D}\equiv \langle \Sigma_{\rm out}\rangle=1,
\end{equation}

where $\{|u_j\rangle\}$ stands for an arbitrary orthonormal basis, with $u_j=|u_j\rangle \langle u_j|$. Note that the unital map need not be a free operation.

Let us illustrate the relevance of this general result for qubits. The crucial point is that for each individual qubit, two orthonormal kets, say $|u_1\rangle$ and $|u_2\rangle$, suffice to construct a basis. In the case of an entangling unital map, at least two qubits must be involved, and the inputs $|u_1\rangle \otimes  |u_1\rangle$ and  $|u_2\rangle \otimes  |u_2\rangle$ do not constitute a basis of the larger Hilbert space. This feature allows the distinction mentioned above. Therefore, instead of a continuum of random input states, one can use $D$ orthonormal states to detect whether a qudit has been acted upon by an entangling unital map or not, in almost all cases.

In the important case $D=2$, one only needs to consider two input states for all qubits.
In section~\ref{Gate ident}, we describe a protocol, based on Proposition 5, to distinguish entangling and non-entangling gates in unknown circuit layers, and provide an experimental proof-of-concept validation using a CNOT layer.

\section{Experimental implementation}

We experimentally investigate QST dynamics using a four-qubit liquid-state NMR processor based on 
${}^{13}$C-labeled \textit{trans}-crotonic acid~\cite{Xinyue_PRL26,Keyi_PRL26,Liu_PRL25,Xinfang_PRL22}. One nuclear spin encodes the system, while the remaining spins implement the ancillary degrees of freedom required for unitary channel dilations. The spin system is effectively initialized in one of the following input states $|0\rangle$, $|+\rangle$, and $|-\rangle$, with $|\pm\rangle=(|0\rangle \pm|1\rangle)/\sqrt{2}$, while the ancillary qubits are prepared in $|0\rangle$. SM~\cite {supp_material} provides further details on molecular parameters, initial-state preparation, and channel implementation.

In all experiments, the central measured quantity is the grand sum, $\Sigma(\varrho)=D\langle f_1|\varrho|f_1\rangle$. For a spin-1/2 system, $\Sigma(\varrho)$ is directly accessible without full quantum state tomography through the transverse magnetization, $\Sigma_{\Gamma}(\varrho)=1+\langle\sigma_x\rangle_{\Gamma}$, where $\langle\sigma_x\rangle_{\Gamma}$ is evaluated on the output state $\varrho_{\Gamma}^{\text{out}}=\Gamma(\varrho^{\text{in}}_{\text{S}})$ after evolution under $\Gamma$. The reduced characterization is not specific to a qubit readout. Rather, it generally follows from Prop. 1.

To test the dual-map description (Prop.~1) experimentally, we consider the phase-damping (PD) channel $\Gamma_{\rm PD}$, with the damping strength $\lambda$ changing across the experiment~(see SM). This channel is unital but not free. For each value of $\lambda$, we independently determine the output grand sum $\Sigma_{\rm PD}(\varrho)$ from the state obtained after implementing the channel and the corresponding dual prediction obtained from the dual evolution of the textureless state  $\Sigma^{\prime}_{\mathrm{PD}}(\varrho)=1+e^{-\lambda/2}\langle\sigma_x\rangle_{\rm in}$, where $\langle\sigma_x\rangle_{\rm in}$ denotes the Pauli expectation value obtained from the input state $\varrho^{\text{in}}_{\text{S}}$. This provides two independent experimental determinations of the same quantity: one from the channel output and the other inferred from the independently measured input state via the dual-map description. 
\begin{figure}[!ht]
\centering
\includegraphics[width=0.95\columnwidth]{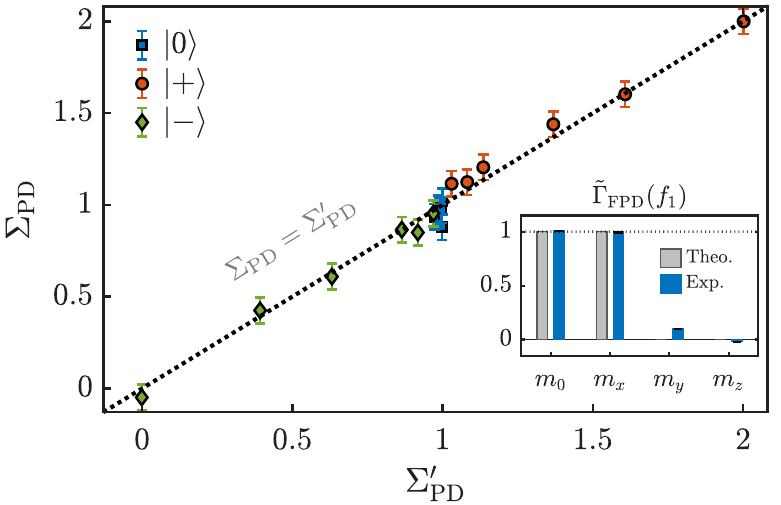}
\caption{Experimental verification of Prop. 1. Comparison between the directly measured output grand sum $\Sigma_{\rm PD}(\varrho)$ and the value $\Sigma^{\prime}_{\rm PD}(\varrho)$ predicted from the dual evolution of the textureless state. Each point corresponds to an independent implementation of the phase-damping channel with a different damping parameter $\lambda$. The diagonal line $\Sigma_{\mathrm{PD}}(\varrho)=\Sigma^{\prime}_{\mathrm{PD}}(\varrho)$ represents the relation predicted by Proposition 1. The agreement between the data and the line confirms that the grand sum response is completely determined by the single operator, $\tilde{\Gamma}_{\rm PD}(f_1)$. Inset: Experimental verification of the dual fixed-point criterion for QST preservation, Prop. 2. The blue bars show the experimentally reconstructed Pauli coefficients of the operator $\tilde{\Gamma}_{\rm FPD}(f_1)$, while the gray bars correspond to the theoretical prediction. The agreement confirms the fixed-point condition, $\tilde{\Gamma}_{\rm FPD}(f_1)=f_1$. Experimental uncertainties stem from the statistical fluctuations in the transverse magnetization and are propagated to the reported quantities using standard error propagation (see SM).}
\label{fig2b}
\end{figure}
 
Each point in Fig.~\ref{fig2b} corresponds to a distinct implementation of the PD channel at a different value of $\lambda$. The diagonal line $\Sigma_{\mathrm{PD}}(\varrho)=\Sigma^{\prime}_{\mathrm{PD}}(\varrho)$ represents the relation predicted by Prop. 1. The agreement of the experimental data with this line provides a direct experimental validation of the dual-map relation for the implemented channel, illustrating the general result established in Prop. 1.

\begin{figure*}[!ht]
\centering
\includegraphics[width=\linewidth]{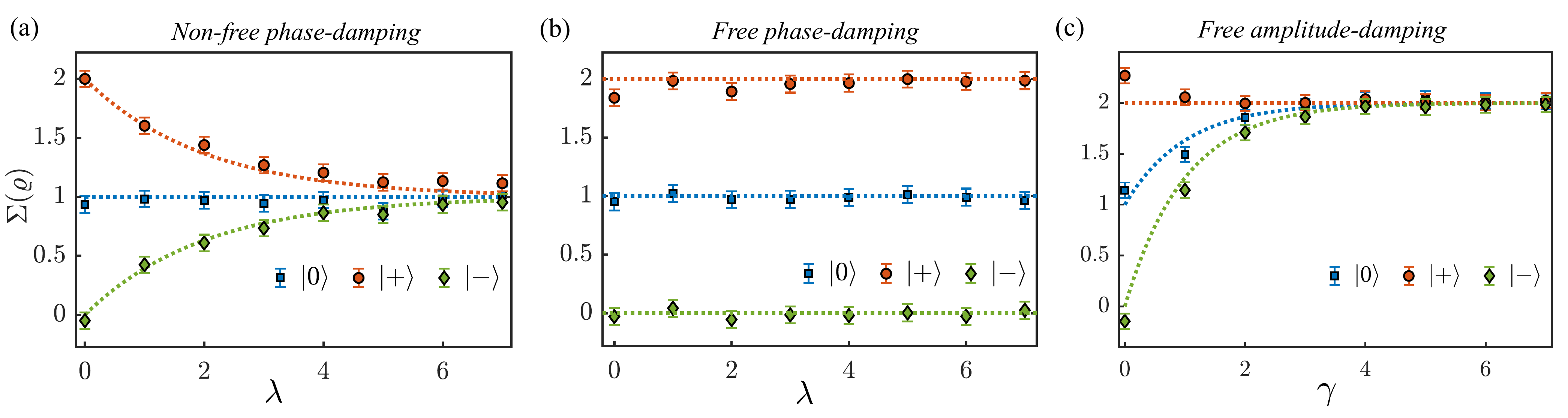}
\caption{Experimental records of QST dynamics. (a) Non-free unital phase-damping channel. The grand sum is redistributed among different input states despite the channel being unital. (b) Free-unital phase-damping channel. The grand sum remains constant throughout the evolution, demonstrating exact QST conservation. (c) Free non-unital amplitude-damping channel. The grand sum evolves toward the textureless state, exemplifying a texture-depleting free map. Symbols denote the experimental data while dotted curves are the theoretical predictions.}
\label{fig03_new}
\end{figure*}

As a second experimental test, we implement the free phase-damping (FPD) channel, which corresponds to dephasing in the Hadamard basis and therefore constitutes a free operation (see SM~\cite{supp_material}). For a spin-1/2 system, the dual action on the textureless state can be written as $\tilde{\Gamma}_{\rm FPD}(f_1)=(m_0\mathds{1}+m_x\sigma_x+m_y\sigma_y+m_z\sigma_z)/2$, where the Bloch coefficients $m_\alpha$ were experimentally reconstructed from the output grand sums measured for a tomographically complete set of input states (see SM). In the inset of Fig.~\ref{fig2b}, we show the agreement between the reconstructed Bloch coefficients (blue bars) and the theoretical prediction. This confirms the fixed-point condition of the dual map $\tilde{\Gamma}_{\mathrm{FPD}}(f_1)=f_1$, and provides an experimental validation of Prop. 2. Importantly, this reconstruction characterizes the full QST response without reconstructing the channel, providing an experimental realization of the reduced process information implied by Prop. 1.

We proceed to investigate representative channels from different sectors of the QST channel hierarchy illustrated in Fig.~\ref{fig1}. Figure~\ref{fig03_new}(a) shows the dynamics under the non-free PD channel, for which the output grand sum is
$\Sigma_{\mathrm{PD}}(\varrho)=1+e^{-\lambda/2}
\left[\Sigma_{\mathrm{in}}(\varrho)-1
\right]$, where $\Sigma_{\mathrm{in}}(\varrho)\equiv\Sigma(\varrho^{\mathrm{in}}_{S})$~(see SM). Although the channel is unital, it does not satisfy the free condition and therefore does not preserve QST. The free-state $|+\rangle$, initially characterized by the maximum grand sum ($\Sigma=2$) and minimal rugosity ($\mathfrak{R}=0$), exhibits a monotonic decrease of $\Sigma$ corresponding to texture generation. Conversely, the maximally textured state $|-\rangle$ shows the opposite behavior, with an increasing grand sum (decreasing texture). Both states evolve toward the balanced value $\Sigma=1$, whereas the intermediate state $|0\rangle$ remains essentially unchanged. These observations illustrate that unitality alone does not imply QST conservation. Instead, general unital dynamics redistributes texture among different input states without preserving their individual QST. This behavior illustrates the balanced redistribution of the grand sum expected for unital dynamics, a property rigorously established in Prop.~4.

Figure~\ref{fig03_new}(b) presents the evolution of the grand sum under the FPD channel, which is both free and unital. As predicted by the dual fixed-point condition [see inset of Fig.~\ref{fig2b}], the grand sum remains constant for every initial state throughout the evolution. 

We next consider a free amplitude-damping (FAD) channel, which is a free but non-unital operation. This channel describes amplitude damping in the Hadamard basis, for which the textureless state $|+\rangle$ turns out to be a fixed point of the dynamics. The resulting grand sum is $\Sigma_{\text{FAD}}(\varrho)=2-[2-\Sigma_{\text{in}}(\varrho)]e^{-\gamma}$ where $\gamma$ is the damping strength. The resulting dynamics are shown in Fig.~\ref{fig03_new}(c). In contrast to the FPD channel [Fig.~\ref{fig03_new}(b)], the FAD channel does not preserve the grand sum of individual input states. While the free state $|+\rangle$ remains invariant throughout the evolution, the states $|0\rangle$ and $|-\rangle$ exhibit a monotonic increase of the grand sum, corresponding to a continuous depletion of their texture. All curves converge to the common value $\Sigma_{\text{FAD}}(\rho)\simeq 2$, in agreement with the theory. 
We plot the corresponding evolution of rugosity under these channels in SM~\cite{supp_material}.

\begin{figure}[!ht]
\centering
\includegraphics[width=0.75\columnwidth]{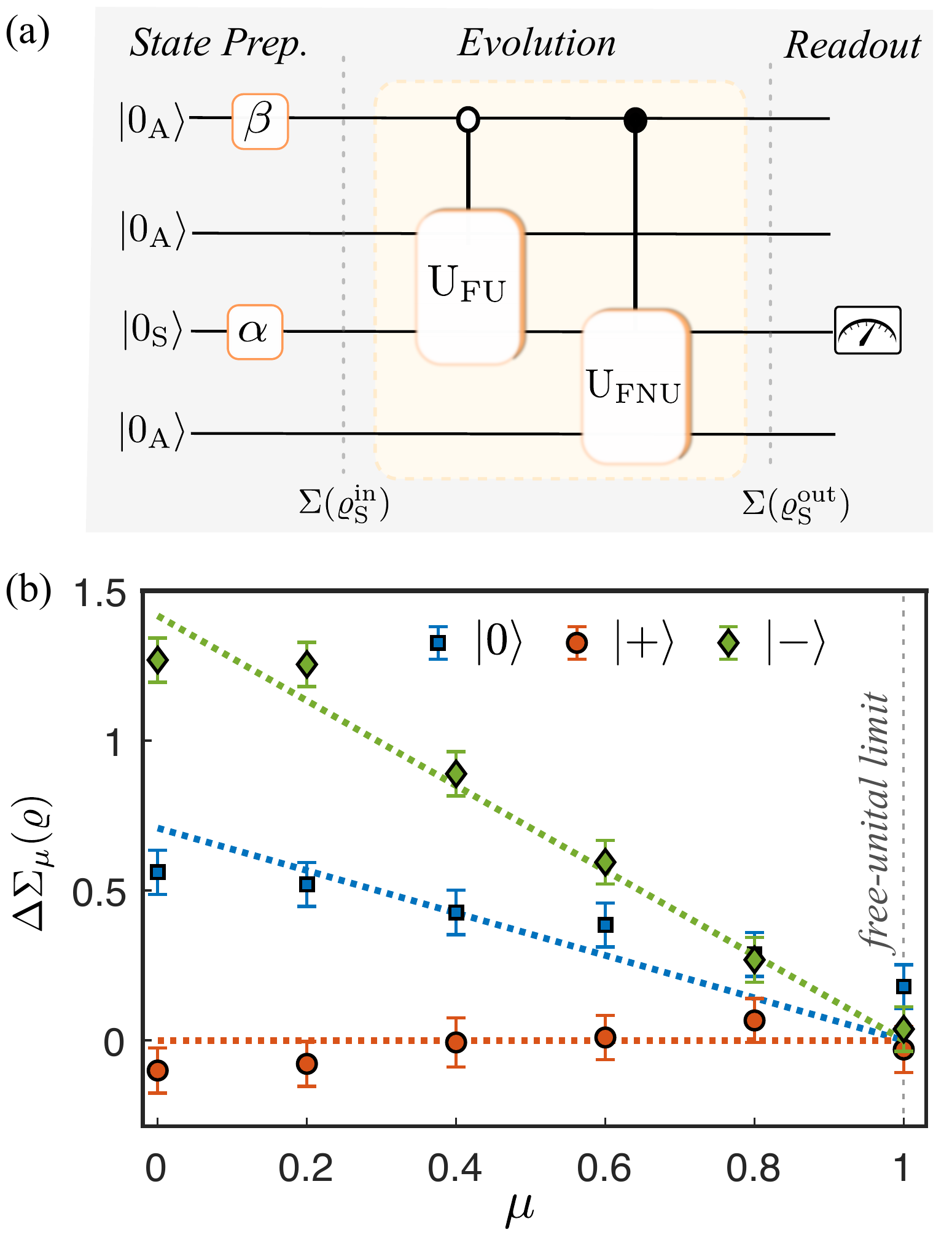}
\caption{Grand sum under composite free dynamics. (a) Quantum circuit implementing a continuous interpolation between free-unital (FU) and free non-unital (FNU) dynamics. The reduced dynamics is described by the map $\Gamma_{\mu}(\varrho)=\mu\Gamma_{\text{FU}}(\varrho)+(1-\mu)\Gamma_{\text{FNU}}(\varrho)$, where $\mu$ is controlled by the ancilla rotation. (b) Measured grand sum variation, $\Delta\Sigma_{\mu}(\varrho)=\Sigma_{\Gamma_{\mu}}(\varrho)-\Sigma_{\text{in}}(\varrho)$ as a function of the control parameter $\mu$, for three input states. The grand sum variation vanishes as the dynamics approach the free-unital limit $\mu=1$, in agreement with Prop.~3. Symbols denote the experimental data, while the dotted lines correspond to the theoretical predictions.}
\label{fig3}
\end{figure}

Now we isolate the role of unitality while remaining entirely within the free sector~(Fig.\ref{fig1}). We investigate how QST evolves under a continuous interpolation between free non-unital (FNU) and free-unital (FU) dynamics, through the four-qubit circuit shown in Fig.~\ref{fig3}(a). To experimentally realize this channel, we introduce a quantum control ancilla prepared in a superposition state, $|\psi_{\beta}\rangle=\cos(\beta/2)|0\rangle_{\rm A}+\sin(\beta/2)|1\rangle_{\rm A}$, that determines which elementary unitary dilation $U_{\rm FU}$ and $U_{\rm FNU}$ is applied. After tracing out the auxiliary qubits, the system dynamics is effectively described by the map $\Gamma_{\mu}(\varrho)=\mu\Gamma_{\text{FU}}(\varrho)+(1-\mu)\Gamma_{\text{FNU}}(\varrho)$, where $\mu=\cos^2(\beta/2)$ is controlled by the rotation angle $\beta$ of the ancilla qubit~(see SM~\cite{supp_material}). Consequently, $\mu$ continuously tunes the dynamics from the FNU to the FU regime. According to Prop.~3, state-independent QST conservation is recovered in the free-unital limit, $\mu=1$. This prediction is experimentally confirmed in Fig.~\ref{fig3}(b), where the grand sum variation $\Delta\Sigma_{\mu}(\varrho)=\Sigma_{\Gamma_{\mu}}(\varrho)-\Sigma_{\text{in}}(\rho)$ decreases monotonically as $\mu$ increases and vanishes, within experimental uncertainty, at $\mu=1$. These results show that, within the present family of free dynamics, state-independent QST conservation is recovered as the non-unital FAD contribution is removed.

The results presented above turn QST into an operational, experimentally accessible quantity (rather than a purely theoretical concept) that can serve as a practical resource for probing quantum circuits. As follows, by exploring local QST measurements, we can distinguish non-entangling unital operations from entangling ones by testing whether the averaged grand sum over an orthonormal basis is preserved after being acted upon by an unknown circuit layer.

\section{Gate identification protocol}\label{Gate ident}

Proposition 5 states that for any non-entangling unital operation, averaging the grand sum over an orthonormal input basis returns 1 for each subsystem. Let $q$ be one of the considered qubits
$\langle\Sigma_{\rm out}\rangle_q=[\Sigma(\varrho_q^{(1)})+\Sigma(\varrho_q^{(2)})]/2=1,$
where the superscript labels $1,2$ refer to two orthogonal input states. An entangling interaction can break this local balance because the inputs do not suffice to form a complete basis of the joint Hilbert space. We exploit this observation to construct a diagnostic based exclusively on local QST measurements.

We first analyze the effect of a CNOT gate whose defining local basis need not coincide with the computational basis. Let $A$ and $B$ denote the control and target qubits, respectively, and let $\{|\downarrow\rangle,\uparrow\rangle\}$, denote the basis in which the CNOT assumes its standard action, 
$U|\uparrow\uparrow\rangle=|\uparrow\downarrow\rangle$, $U|\uparrow\downarrow\rangle=|\uparrow\uparrow\rangle$,
$U|\downarrow\uparrow\rangle=|\downarrow \uparrow\rangle$, $U|\downarrow\downarrow\rangle=|\downarrow\downarrow\rangle.$

The protocol proceeds as follows. To interrogate the layer, we first initialize the two qubits in the product state $|11\rangle$, where $|1\rangle=\alpha|\uparrow\rangle+\beta|\downarrow\rangle$, with $\alpha$ and $\beta$ unknown, encoding the relative orientation between the probing basis and the local basis defining the CNOT. The grand sum related to the reduced density matrix of qubit $A$, after the action of the CNOT, becomes
\begin{equation*}
\Sigma(\varrho_A^{(1)})=1+(|\alpha|^2-|\beta|^2)(\alpha \beta+\alpha^* \beta^*)(1-\alpha^* \beta-\alpha \beta^*).
\end{equation*}
In the second preparation, the two qubits are initialized in the product state $|22\rangle$, where $|2\rangle=\beta^*|\uparrow\rangle-\alpha^*|\downarrow\rangle$. Similar calculations lead to:
\begin{equation*}
\Sigma(\varrho_A^{(2)})=1+(|\alpha|^2-|\beta|^2)(\alpha \beta+\alpha^* \beta^*)(1+\alpha^* \beta+\alpha \beta^*).
\end{equation*}
Averaging over the two orthogonal preparations yields
\begin{eqnarray}
\langle \Sigma_{\rm out}\rangle_A=1+2(|\alpha|^2-|\beta|^2)\Re(\alpha \beta).
\end{eqnarray}
Since this quantity is different from unity, except for $|\alpha|=|\beta|$ and/or $\Re(\alpha \beta)=0$, one can infer that qubit $A$ participated in an entangling operation.

Performing the analogous calculation for qubit $B$ gives
\begin{eqnarray}
\langle \Sigma_{\rm out}\rangle_B=1-2[\Re(\alpha^2)-\Re(\beta^2)]\Re(\alpha^* \beta).
\end{eqnarray}
Again, this quantity is different from unity, except when $\Re(\alpha^2)=\Re(\beta^2)$, and/or $\Re(\alpha^* \beta)=0$.  Consequently, except for a zero-measure set, at least one of the two local averages deviates from unity, providing a signature of the CNOT interaction, for example, the choice  $\alpha=1/\sqrt{2}$ and $\beta=-i/\sqrt{2}$.
\begin{figure}[!ht]
\centering
\includegraphics[width=\columnwidth]{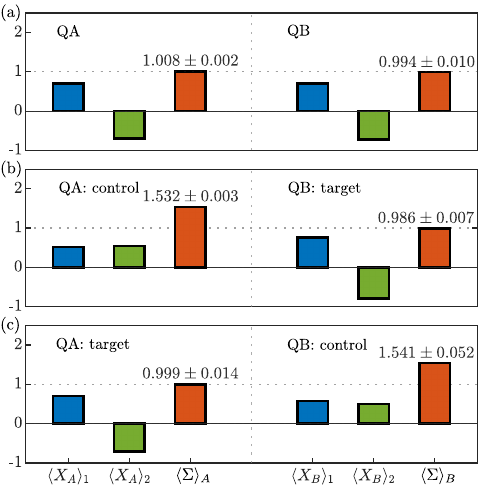}
\caption{QST-based diagnosis of a two-qubit entangling layer. All experiments are performed in the four-qubit register, while only the local responses of the interacting pair $A, B$ are displayed. (a) Reference experiment containing only local unitary operations illustrating that the grand sum average (red bars) over the two orthogonal input preparations remains balanced. Here, $\langle X_{A}\rangle_{1,2}$ and $\langle X_{B}\rangle_{1,2}$ denotes the $x$-magnetization measured for the two input states. (b,c) A CNOT breaks the local balance. Experimental identification of the control and target qubits for the $\mathrm{CNOT}_{A\rightarrow B}$ and $\mathrm{CNOT}_{B\rightarrow A}$ gates, respectively. The observed imbalance follows the gate orientation, providing a local signature of the directed entangling interaction. Bars denote experimental data and horizontal dashed lines denote the theoretical predictions.}
\label{fig4}
\end{figure}

We now experimentally test the layer-resolved diagnostic protocol using a four-qubit NMR register in two complementary scenarios. In both experiments, the circuit is interrogated through the same two orthogonal product-state preparations $|1\rangle^{\otimes4}$ and $|2\rangle^{\otimes4}$, with $|1\rangle=\cos(\theta/2)|0\rangle+\sin(\theta/2)|1\rangle$,$|2\rangle=\sin(\theta/2)|0\rangle-\cos(\theta/2)|1\rangle$, and $\theta=\pi/4$. For each qubit $q$, the diagnostic quantity is the basis-averaged local grand sum $\langle\Sigma_{\rm out}\rangle_q=1+(\langle X_{q}\rangle_{1}+\langle X_{q}\rangle_{2})/2$, where $\langle X_{q}\rangle_{1,2}$ corresponds to the $x$-component of the transverse magnetization for each qubit $q\in\{A,B,C,D\}$ for the input states. We stress that the goal is not to reconstruct the implemented circuit layer, but to determine from these local quantities whether the layer is compatible with local operations or contains an entangling interaction. 

We first consider a benchmark scenario in which a CNOT, written in the computational basis, acts upon the selected pair $A,B$, while qubits $C,D$ are spectators. In the chosen probing basis, the two CNOT orientations yield distinct theoretical local-QST fingerprints, allowing us to resolve the control-target orientation. In what follows, we compare two situations: an evolution in which we apply local (non-entangling) unitary rotations, $U=R^{(A)}_x(\pi/3)\otimes R^{(B)}_x(2\pi/5)\otimes \mathds{1}^{(C)}\otimes \mathds{1}^{(D)}$, with an evolution that includes a CNOT gate between the pair of qubits $A, B$. Figure~\ref{fig4}(a) shows the measured average grand sums under local gates. In this case, the averaged grand sums remain around the balanced values $[\langle\Sigma\rangle_{A},\langle\Sigma\rangle_{ B},\langle\Sigma\rangle_{C},\langle\Sigma\rangle_{ D}]\simeq[1,1,1,1]$, in agreement with Prop. 5. Since this first experiment aims to resolve the interacting pair's response and its control-target orientation, Fig.~\ref{fig4}(a) displays the local observables associated only with the qubits $A$ and $B$.

Conversely, the CNOT gate breaks local balancedness. For $\mathrm{CNOT}_{{q_{c}}\rightarrow {q_{t}}}$, where $q_c$ and $q_t$ denote the control and target qubits, respectively, the ideal averaged grand sums are $[\langle\Sigma\rangle_{q_{c}},\langle\Sigma\rangle_{q_{t}}]=[1.5,1]$. By reversing the gate orientation, $\mathrm{CNOT}_{{q_{t}}\rightarrow {q_{c}}}$, these values are interchanged, and thus we can distinguish the two possible gate orientations. The experimental results shown in Figs.~\ref{fig4}(b) and \ref{fig4}(c) reproduce these predictions, demonstrating that the local QST imbalance provides a diagnostic signature of the entangling layer and resolves its control-target orientation.

In this benchmark, we assume that the basis in which the CNOT is implemented is known when interpreting the results. Next we test a more stringent scenario in which this information is deliberately excluded from the identification procedure. More precisely, we implement the physical gate in a rotated local basis chosen during the experiment, but the diagnostic receives only the two fixed input preparations and the measured local grand sums. It is therefore necessary to identify the qubits participating in the interaction without using the basis transformation that defines the CNOT.
\begin{figure}[!ht]
\centering
\includegraphics[width=\linewidth]{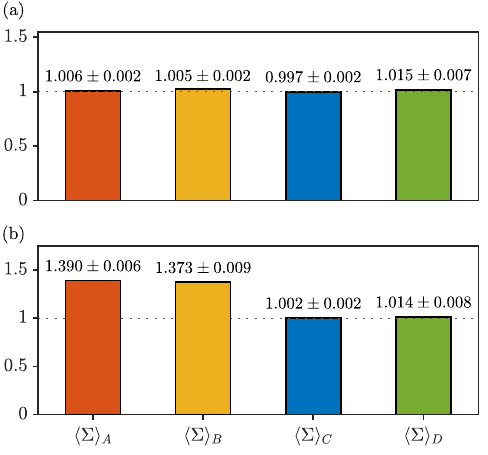}
\caption{QST-based localization of an entangling interaction in a rotated local basis. The bars show the basis-averaged local grand sum $\langle\Sigma_{\rm out}\rangle_q=1+(\langle X_{q}\rangle_{1}+\langle X_{q}\rangle_{2})/2$ measured for each individual qubit $q=\{A,B,C,D\}$. (a) A layer composed exclusively of local unitary rotations preserves basis balancedness for all four qubits. (b) A CNOT acting on qubits $A$ and $B$ is physically implemented in the rotated basis defined by $V=R_z(-\pi/2)R_y(\pi/4)$, while this transformation is not used by the identification procedure. The deviations from unity on $A$ and $B$ identify them as the qubits participating in the entangling interaction, whereas spectator qubits $C$ and $D$ remain balanced. The dotted line marks the theoretical predictions for $\langle\Sigma\rangle_{q}=1$.}
\label{fig4b}
\end{figure}
Figure~\ref{fig4b}(a) establishes the local reference for this second test. In this case, we consider the non-entangling circuit layer composed of different local rotations, $U_{\rm loc}=R_y^{(A)}(\pi/3)\otimes R_z^{(B)}(\pi/3)
\otimes R_y^{(C)}(-\pi/3)
\otimes[
R_z^{(D)}(\pi/4)R_y^{(D)}(\pi/2)]$. Although the individual values of the grand sum depend on the applied rotation and assume different values for each qubit, Prop. 5 ensures that their averages over the two orthogonal preparations remain balanced. Experimentally, all four qubits satisfy $\langle\Sigma_{q}\rangle\simeq 1$, within experimental uncertainties. We then implement a CNOT between qubits $A$ and $B$ in the rotated local basis, $$|\downarrow\rangle=V|0\rangle,\quad |\uparrow\rangle=V|1\rangle,\quad V=R_z(-\pi/2)R_y(\pi/4),$$ corresponding to the non-local operation $U_{AB}^{(V)}=(V\otimes V)\text{CNOT}_{A\rightarrow B}(V^{\dag}\otimes V^{\dag})$. Importantly, the transformation $V$ is used only to implement the CNOT gate in the new basis, but is not used in the identification step. All the diagnostics are based solely on the measured values of $\langle\Sigma_{q}\rangle$ for each qubit.

Figure~\ref{fig4b}(b) shows that the interaction produces local QST imbalances on the participating qubits, $[\langle\Sigma\rangle_{A},\langle\Sigma\rangle_{ B}]\simeq[1.390,1.373],$ while the spectator qubits remain compatible with the balanced baseline $[\langle\Sigma\rangle_{C},\langle\Sigma\rangle_{D}]\simeq[1.002, 1.014].$ Thus, from the local QST data alone, the qubits $A$ and $B$ are identified as the subsystems participating in the entangling layer, even though the local basis defining the CNOT is not provided for the diagnostic. This demonstrates localization of an entangling interaction without prior knowledge of the local basis in which the CNOT is implemented. Taken together, Fig.~\ref{fig4} and Fig.~\ref{fig4b} illustrate two complementary capabilities of the protocol: the controlled benchmark resolves the orientation of a CNOT acting on a selected pair, while the hidden-basis test localizes the interacting pair within the full register using only local QST measurements.

A final comment is in order. One might argue that the previous results apply only to detecting CNOT gates, not general entangling gates. However, any two-qubit entangling gate can be decomposed into single-qubit unitary operations plus a finite number of CNOT gates. In addition, conventional characterization of unknown gates typically relies on quantum process tomography (QPT), which requires reconstructing the implemented operation. In contrast, the present QST-based protocol identifies entangling operations without reconstructing the full process. All information comes solely from local measurements, reducing the experimental overhead compared with full QPT. Beyond this proof-of-concept demonstration, our protocol also suggests a route toward layer-resolved diagnosis of circuit connectivity in larger quantum registers.

\section{CONCLUSION}

In this work, we addressed two outstanding questions, namely, (i) the theoretical structure underlying QST dynamics and (ii) the practical demonstration that QST is experimentally accessible. We developed a general framework describing the dynamics of quantum-state texture under arbitrary finite-dimensional CPTP maps and demonstrated that the transformed texture of any state is fully determined by the action of the corresponding dual map on the textureless state. We found necessary and sufficient conditions for QST conservation and identified free-unital maps as an important class of protective channels. We experimentally tested these predictions using representative quantum channels and further demonstrated the continuous recovery of state-independent QST conservation as a family of free dynamics approaches the unital regime. Finally, we showed that the developed framework naturally enables operational applications and is experimentally accessible. In particular, local QST can provide a layer-resolved signature of entangling interactions without reconstructing the full quantum process. Beyond this proof-of-concept demonstration, this approach suggests QST can be used to diagnose pairwise local connectivity in larger quantum circuits. These results establish quantum-state texture as an accessible resource and useful diagnostic for quantum information processes.

\begin{acknowledgments}
C.H.S.V. acknowledges the Funda\c{c}\~ao de Amparo \`a Pesquisa do Estado de S\~ao Paulo (FAPESP - Grant No. 2023/13362-0 and 2025/14546-2) for financial support and the Southern University of Science and Technology (SUSTech) for providing the workspace during the research internship. X.N. and D.L. are supported by the National Natural Science Foundation of China (Grants No. 12574543, 12575020), Guangdong Provincial Quantum Science Strategic Initiative (GDZX2203001, GDZX2303001,  GDZX2403004, GDZX2503001, GDZX2506002), and Guangdong Basic and Applied Basic Research Foundation (2025A1515011599, 2026B1515020012). F. P. acknowledges financial support from the Brazilian agencies Coordena\c{c}\~ao de Aperfei\c{c}oamento de Pessoal de N\'{\i}vel Superior (CAPES), Conselho Nacional de Desenvolvimento Cient\'{\i}fico e Tecnol\'ogico through its program CNPq INCT-IQ (Grant 465469/2014-0), Funda\c{c}\~ao de Amparo \`a Pesquisa do Estado de S\~ao Paulo (FAPESP - Grant 2021/06535-0), and Funda\c{c}\~ao de Amparo \`a Ci\^encia e Tecnologia do Estado de Pernambuco (FACEPE - Grant BPP-0037-1.05/24). 

\end{acknowledgments}

\vspace{0.5cm}
\textit{Data availability.}
All relevant experimental data are presented in the manuscript. Additional data generated or analyzed during this work are available from the corresponding author upon reasonable request.




\bibliography{refs}

\newpage
\pagebreak
\clearpage
\onecolumngrid

\begin{center}
\vskip0.5cm
{\Large \textbf{Supplemental Material: Quantum-State Texture Dynamics: Theory and Experiment}}

\vskip0.6cm
{
Carlos H. S. Vieira$^{1,2}$, Xinfang Nie$^{3,2}$, Dawei Lu$^{2,3}$ and Fernando Parisio$^{4}$
}

\vskip0.2cm
{\small
$^{1}$~\textit{Centro de Ci\^{e}ncias Naturais e Humanas, Universidade Federal do ABC,
Avenida dos Estados 5001, 09210-580 Santo Andr\'e, S\~{a}o Paulo, Brazil}\\
$^{2}$~~\textit{Department of Physics, State Key Laboratory of Quantum Functional Materials,
and Guangdong Basic Research Center of Excellence for Quantum Science, Southern University of Science and Technology, Shenzhen 518055, China} \\
$^{3}$~\textit{Quantum Science Center of Guangdong-HongKong-Macao Greater Bay Area, Shenzhen 518045, China}\\
$^{4}$~\textit{Departamento de F\'{\i}sica, Centro de Ci\^encias Exatas e da Natureza, Universidade Federal de Pernambuco, Recife, Pernambuco
50670-901 Brazil}
}

\end{center}
\vskip0.4cm

\onecolumngrid

\author{Xinfang Nie\,\orcidlink{0000-0000-0000-0000}}
\affiliation{Department of Physics, State Key Laboratory of Quantum Functional Materials,
and Guangdong Basic Research Center of Excellence for Quantum Science,
Southern University of Science and Technology, Shenzhen 518055, China}
\affiliation{Quantum Science Center of Guangdong-HongKong-Macao Greater Bay Area, Shenzhen 518045, China}


\setcounter{equation}{0}
 \setcounter{figure}{0}
 \setcounter{table}{0}
 \setcounter{page}{1}
 \renewcommand{\theequation}{S\arabic{equation}}
 \renewcommand{\thefigure}{S\arabic{figure}}

\renewcommand{\thepage}{S\arabic{page}}

\thispagestyle{empty}  
\pagestyle{fancy}
\fancyhf{}
\fancyfoot[C]{\thepage}
\renewcommand{\headrulewidth}{0pt}
\renewcommand{\footrulewidth}{0pt}

\renewcommand{\thepage}{S\arabic{page}}
\setcounter{page}{1}

This Supplemental Material provides detailed proofs and experimental information supporting the main text. It includes derivations of the QST dynamical results, additional discussion of representative quantum channels, and details of the experimental implementations.

\section{Proofs of propositions}
In this section, we provide detailed proofs of the main theoretical propositions and comment on some relevant consequences. We denote arbitrary CPTP maps by $\Gamma$ (effected by Kraus operators $K_j$) and free channels by $\Lambda$ (effected by Kraus operators $E_j$).

{\bf Proposition 1}: Let $\varrho \in B({\cal H})$ be an arbitrary finite-dimensional quantum state and $\Gamma$ a general CPTP map effected by the Kraus operators $\{K_j\}$. Then, the grand sum of $\Gamma(\varrho)=\sum_jK_j \varrho K_j^{\dagger}$ is given by:
\begin{equation}
\label{generalSM}
\Sigma(\Gamma(\varrho))=D {\rm Tr}(\tilde{\Gamma}_1 \varrho) =D \Sigma(\tilde{\Gamma}_1^{\rm T} \odot \varrho),
\end{equation}
where $D$ is the dimension of ${\cal H}$, $\tilde{\Gamma}(f_1)\equiv \tilde{\Gamma}_1= \sum_jK^{\dagger}_j f_1 K_j$, T denotes transposition and $\odot$ denotes the Hadamard (or entrywise) product [given two matrices with the same dimension $A$ and $B$, then $(A \odot B)_{ij}\equiv A_{ij}B_{ij}$]. We provide two different demonstrations of Prop. 1. The first proof is shorter, and the second is more elementary. 

{\it Proof A:} By definition, the dual map $\tilde{\Gamma}$ to a channel $\Gamma$ is such that 
${\rm Tr}(\sigma\Gamma(\varrho))={\rm Tr}(\varrho\tilde{\Gamma}(\sigma))$ for all pairs $\varrho$ and $\sigma$ in the Hilbert-Schmidt space. Since the grand sum of $\Gamma(\varrho)$ can be written as $\Sigma(\Gamma(\varrho))=D{\rm Tr}(f_1\Gamma(\varrho))$, then $\Sigma(\Gamma(\varrho))=D{\rm Tr}(\tilde{\Gamma}(f_1)\varrho)$, which justifies the first equality in (\ref{generalSM}). Therefore, 
$$D {\rm Tr}(\tilde{\Gamma}_1 \varrho)=D\sum_n\langle n|\tilde{\Gamma}_1\varrho |n\rangle=D\sum_{n,j}\langle n|\tilde{\Gamma}_1|j\rangle \langle j|\varrho |n\rangle=D \Sigma(\tilde{\Gamma}_1^{\rm T} \odot \varrho),$$ 

which proves the second equality in (\ref{generalSM}). $\blacksquare$

{\it Proof B:} We have
$$\Sigma(\Gamma(\varrho))=D\langle f_1|\Gamma(\varrho)|f_1\rangle=D\sum_j\langle f_1|K_j\varrho K_j^{\dagger}|f_1\rangle=D\sum_j\langle f_1|K_j\varrho K_j^{\dagger}|f_1\rangle=D\sum_j\sum_{\ell,n}\langle f_1|K_j|u_{\ell}\rangle\varrho_{\ell n}\langle u_n| K_j^{\dagger}|f_1\rangle,$$
where $\{|u_{\ell}\rangle\}$ is an arbitrary orthonormal basis. Therefore, we can write
$$\Sigma(\Gamma(\varrho))=D\sum_{\ell,n}\langle u_n|\left(\sum_j K_j^{\dagger}f_1 K_j\right)|u_{\ell}\rangle\varrho_{\ell n}=D\sum_{\ell,n}[\tilde{\Gamma}_1]_{n \ell}\varrho_{\ell n}=D\sum_{\ell,n}[\tilde{\Gamma}_1^{\rm T}]_{\ell n}\varrho_{\ell n}=D\sum_{\ell,n}[\tilde{\Gamma}_1^{\rm T} \odot \varrho]_{\ell n}=D\Sigma(\tilde{\Gamma}_1^{\rm T} \odot \varrho). \blacksquare$$

In the operator-sum representation, the dual map is expressed by $\tilde{\Gamma}(\cdot)=\sum_jK^{\dagger}_j (\cdot) K_j$, and thus $\tilde{\Gamma}_1=\sum_jK_j^{\dagger} f_1 K_j$.
Therefore, to calculate the transformed grand sum of any $\varrho$, one just needs to compute how the single state $f_1$ transforms under the dual map. 

Before we proceed, let us establish some key properties of $\tilde{\Gamma}_1$. First, it is clear from its definition that $\tilde{\Gamma}_1=\tilde{\Gamma}_1^{\dagger}$. That $\tilde{\Gamma}$ is completely positive follows from the operator sum representation through which it is defined.
In fact, for an arbitrary pure state $|\psi\rangle$, we have $\langle \psi|\tilde{\Gamma}_1|\psi\rangle=\sum_j  \langle f_1|K_j|\psi\rangle\langle \psi| K_j^{\dagger}|f_1\rangle= \langle f_1|\Gamma(\psi)|f_1\rangle\ge 0$.
Note also that since $0 \le \Sigma(\Gamma(\varrho))\le D$, we must have $0\le  \Sigma(\tilde{\Gamma}_1^{\rm T} \odot \varrho) \le 1$.

In addition, whenever $\Gamma$ is a unital map, $\Gamma(\mathds{1})=\mathds{1}$, $\tilde{\Gamma}_1$ is a density matrix. Given the previous properties, we only have to show that ${\rm Tr}(\tilde{\Gamma}_1)=1$. Indeed ${\rm Tr}(\tilde{\Gamma}_1)=\sum_{n,j}\langle n|K_j^{\dagger}|f_1\rangle \langle f_1|K_j|n \rangle=\sum_{j}\langle f_1|K_jK_j^{\dagger}|f_1\rangle$. Using the unitality condition, we get ${\rm Tr}(\tilde{\Gamma}_1)=1$.

{\bf Proposition 2}:  A general channel $\Lambda$ is QST-preserving if and only if $\tilde{\Lambda}_1=f_1$, that is, $f_1$ is a fixed point of the dual map $\tilde{\Lambda}$. Also, for any QST-preserving map, all Kraus operators commute with $f_1$.

{\it Proof:} From the second equality in Eq. (\ref{generalSM}), the QST preservation reads $D \Sigma(\tilde{\Lambda}_1^{\rm T} \odot \varrho)= \Sigma(\varrho)$, or $D \Sigma(\tilde{\Lambda}_1^{\rm T} \odot \varrho-\varrho/D)=0$. But, in general, $\varrho/D=f_1 \odot \varrho$, for any $\varrho$ (for the Hadamard product $Df_1$ functions as the identity operator). The preservation condition becomes $$\Sigma\left( (\tilde{\Lambda}_1^{\rm T}-f_1)\odot \varrho \right)=0,$$ for all $\varrho$. This condition is generally satisfied if $\tilde{\Lambda}_1^{\rm T}-f_1=0$, or $\tilde{\Lambda}_1=f_1^{\rm T}$. But $f_1^{\rm T}=f_1$.

Now we prove the second statement.  We assume that the map is QST-preserving and show that $[f_1, E_j ] = 0$. If the channel is QST-preserving, then $\Lambda(f_1)=f_1$ and $\tilde{\Lambda}(f_1)=f_1$. Since $f_1$ is a pure state (an extremal point in the space of states), these relations imply $E_j|f_1\rangle=\alpha_j|f_1\rangle$ and $E_j^{\dagger}|f_1\rangle=\beta_j|f_1\rangle$, respectively. This immediately implies $\langle f_1|E_j|f_1\rangle=\alpha_j$ and $\langle f_1|E_j^{\dagger}|f_1\rangle=\beta_j$ and thus $\beta_j=\alpha_j^{*}$. Therefore, one can write $E_jf_1=\alpha_j f_1$ and 
\begin{equation}
f_1E_j=[( f_1E_j)^{\dagger}]^{\dagger}=(E_j^{\dagger} f_1)^{\dagger}=(\alpha_j^* f_1)^{\dagger}=\alpha_jf_1,
\end{equation}
where we used the Hermiticity of $f_1$. Therefore, $[f_1,E_j]=0$. $\blacksquare$

{\bf Proposition 3}: Any free unital CPTP map necessarily preserves quantum-state texture. 

{\it Proof:} This is a direct consequence of a previous theorem by Watrous \cite{watrous} valid for finite-dimensional systems. It states that, a unital CPTP map $\Lambda$ has $\varrho$ as a fixed point, $\Lambda(\varrho)=\sum_jE_j\varrho E_j^{\dagger}=\varrho$  if and only if $[\varrho,E_j]=0$, for all $j$. Since the channel is free, we have $\Lambda(f_1)=f_1$ and thus $[f_1,E_j]=0$, which also implies $[f_1,E_j^{\dagger}]=0$. We have $$\tilde{\Lambda}(f_1)=\sum_jE_j^{\dagger}f_1E_j=f_1\sum_jE_j^{\dagger}E_j=f_1.$$ This last equality is the QST-preserving condition.  $\blacksquare$

{\bf Proposition 3b}: For a two-dimensional Hilbert space, a free CPTP map $\Lambda$ preserves quantum-state texture if and only if it is unital.

{\it Proof:} The implication from free unitality to QST preservation follows from Prop. 3 in arbitrary finite dimension. It therefore remains only to prove the converse for $D=2$.

Let $\Lambda(\varrho)=\sum_{j}E_{j}\varrho E^{\dagger}_{j}$ be a free CPTP map acting on a two-dimensional Hilbert space, and let $\{|f_1\rangle, |f_2\rangle\}$ be an orthonormal basis, where $f_1=|f_1\rangle\langle f_1|$ is the textureless state. Because $f_1$ has rank one and each operator $E_j f_1 E_j^\dagger$ is positive semidefinite, every vector $E_j|f_1\rangle$ must belong to the one-dimensional support of $f_1$, and this implies that $E_{j}|f_1\rangle=\alpha_{j}|f_1\rangle$, for every $j$. Hence, in the basis $\{|f_1\rangle,|f_2\rangle\}$, each Kraus operator has the form $E_{j}=[\alpha_j, b_{j}; 0, d_{j}]$. If the channel preserves QST, Proposition 2 implies $\tilde{\Lambda}(f_1)=\sum_j E_j^\dagger f_1 E_j=f_1$. Applying the same rank-one argument to this relation yields
$E^{\dagger}_{j}|f_1\rangle=\beta_{j}|f_1\rangle$. Since $E^{\dagger}_{j}=[\alpha^{\ast}_{j}, 0; b^{\ast}_{j}, d^{\ast}_{j}]$, we have that $b_{j}=0$. Thus, every Kraus operator is diagonal in this basis $E_{j}=[\alpha_{j}, 0; 0, d_{j}]$.

Trace preservation requires, $\sum_{j}E^{\dagger}_jE_j=\mathds{1}$, which gives $\sum_{j}|\alpha_j|^2=1$ and $\sum_{j}|d_j|^2=1$. Since each $E_j$ is diagonal, $E_jE^{\dagger}_j=E^{\dagger}_jE_j$. Therefore, $\sum_jE_jE_j^\dagger=\sum_jE_j^\dagger E_j=\mathds{1}$, and hence $$\Lambda(\mathds{1})=\mathds{1}.$$ Thus, any QST-preserving free CPTP map is necessarily unital for $D=2$. $\blacksquare$ 

{\it Remark.} This equivalence is specific to qubits because the subspace orthogonal to the textureless state is one-dimensional. Consequently, the free and QST-preserving conditions force each Kraus operator to decouple $|f_1\rangle$ from its orthogonal complement, while the remaining block acting on that complement reduces to a complex scalar. In this case, trace preservation also ensures unitality. For $D\ge3$, however, the orthogonal complement has dimension at least two, so the corresponding Kraus blocks are nontrivial matrices and may support non-unital dynamics entirely within this subspace, while preserving the overlap with $|f_1\rangle$ and, consequently, the QST.

{\bf Proposition 4} (Haar balancedness): An arbitrary unital map,  $\Gamma$, preserves the average grand sum with respect to the Haar measure. That is,
\begin{equation*}
\overline{\Sigma}_{\rm in}\equiv \int d\Omega \Sigma({\psi_{\rm in}})  = \int d\Omega \,\Sigma(\Gamma({\psi_{\rm in}}))\equiv\overline{\Sigma}_{\rm out}=1,
\end{equation*}
where $ d\Omega$ represents the Haar integration measure. 

{\it Proof:} Let us initially consider pure arbitrary initial qudit states, $\psi_{\rm in}=|\psi_{\rm in}\rangle \langle \psi_{\rm in}|$. We have
\begin{equation*}
\overline{\Sigma}_{\rm in}=\int d\Omega  D{\rm Tr}\left( f_1 \psi_{\rm in}  \right)= D {\rm Tr}\left( f_1\int d\Omega \psi_{\rm in}\right)=1,
\end{equation*}
where we used $\int d\Omega \, \psi_{\rm in}=\mathds{1}/D$.
We now show that this average is unchanged under an arbitrary single-qudit unital operation. Let  $ \psi_{\rm out}\equiv\Gamma(\psi_{\rm in})=\sum_j K_j \psi_{\rm in} K_j^{\dagger} $. In this case, the output average reads
\begin{eqnarray}
\nonumber
\overline{\Sigma}_{\rm out}&=&\int d\Omega \, D{\rm Tr}\left(f_1  \Gamma( \psi_{\rm in})\right)
= D {\rm Tr}\left( f_1\int d\Omega \, \Gamma( \psi_{\rm in})\right)\\
\nonumber
&=&D {\rm Tr}\left( f_1\sum_jK_j\left( \int d\Omega \, \psi_{\rm in}\right)K_j^{\dagger}\right)\\
&=&D {\rm Tr}\left( f_1\Gamma(\mathds{1}/D)\right)= {\rm Tr}( f_1)=1,
\end{eqnarray}
where we used the unitality condition in the last line.  $\blacksquare$

{\bf Proposition 5} (Basis balancedness):  An arbitrary unital map,  $\Gamma$, preserves the average grand sum over any orthonormal basis. That is,
\begin{equation*}
\langle \Sigma_{\rm in} \rangle \equiv\frac{1}{D}\sum_{j=1}^D\Sigma(u_j)=\frac{1}{D}\sum_{j=1}^D\Sigma(\Gamma(u_j))\equiv \langle \Sigma_{\rm out}\rangle=1,
\end{equation*}
where $\{|u_j\rangle\}$ stands for an arbitrary orthonormal basis, and $u_j=|u_j\rangle \langle u_j|$. 

{\it Proof:} It is evident that $\langle \Sigma_{\rm in} \rangle=\sum_j\langle f_1|u_j|f_1\rangle=\langle f_1|\sum_ju_j|f_1\rangle=1$.
Consider that in the operator-sum representation we have $\Gamma(\varrho)=\sum_nK_n\varrho K_n^{\dagger}$ and, thus, $\Gamma(u_j)=\sum_nK_nu_j K_n^{\dagger}$. 
Therefore, $$\sum_j\Gamma(u_j)=\sum_nK_n\left(\sum_ju_j\right) K_n^{\dagger}=\sum_nK_n K_n^{\dagger}.$$ From the unitality condition $\sum_j\Gamma(u_j)=\mathds{1}$ and 
$\sum_j\langle f_1|\Gamma(u_j)|f_1\rangle=1$. $\blacksquare$

\subsection{Magic unitaries and texture-destroying channel}

For a general closed system, all physical operations on an arbitrary state $\varrho$ can be expressed through some unitary operator, that is, $\Gamma(\varrho)=U\varrho U^{\dagger}$. Unitaries are extremal maps (they cannot be expressed as non-trivial convex combinations of other channels). In the case of QST, free unitaries must satisfy $Uf_1U^{\dagger}=f_1$.

We now characterize the free unitaries in terms of the so-called magic matrices (do not confuse with the terminology ``magic'' often used in the context of non-stabilizerness). These matrices are such that all their rows and all their columns sum to the same constant value: $\sum_i U_{ij}=\sum_iU_{ji}=const$, for all $j$.

{\bf Proposition 6}: A unitary $U$ is a free operation if and only if it is a magic unitary, satisfying $\sum_i U_{ij}=\sum_iU_{ji}=e^{i\varphi}$, for all $j$ and $\varphi \in[0,2\pi)$.

{\it Proof:} The unitary $U$, with matrix elements $U_{ij}$ in the selected basis is free if $Uf_1U^{\dagger}=f_1$, that is, $U|f_1\rangle=e^{i\varphi}|f_1\rangle$ which is equivalent to $\sum_j U_{ij}[f_1]_{j}=e^{i\varphi}[f_1]_{i}$. But $[f_1]_{i}=1/\sqrt{D}$ for all $i$, and thus 
$\sum_j U_{ij}=e^{i\varphi}$. One can similarly show $\sum_j U_{ji}=e^{i\varphi}$ using the fixed-point conditions for bras. Therefore, if $U$ is a free operation, we must have $\sum_i U_{ij}=\sum_iU_{ji}=e^{i\varphi}$, for all $j$, which is the definition of magic unitaries whose columns and rows sum to 1 (up to an irrelevant global phase). Again, using $[f_1]_{i}=1/\sqrt{D}$ and assuming that the unitary is magic, we get 
$$\sum_j U_{ij}[f_1]_{j}=\frac{1}{\sqrt{D}}\sum_j U_{ij}=\frac{e^{i\varphi}}{\sqrt{D}}=[v]_{j}.$$ Therefore, $[v]_{j}=e^{i\varphi}[f_1]_{j}$. $\blacksquare$

The previous channels constitute simple instances of general unital maps. 
As a relevant example of a free non-unital channel, we address the resource-destroying map. This class of maps does not necessarily exist in general resource theories, as, for instance, in entanglement theory. A resource-destroying map $\Delta$ must satisfy the following two properties \cite{gour}: (i) If $\varrho$ is a free state, then $\Delta(\varrho)=\varrho$ (invariance) and (ii) For an arbitrary $\varrho$, $\Delta(\varrho)$ is necessarily a free state (destruction). 

{\bf Proposition 7}: QST resource theory admits a texture-destroying map.

{\it Proof:} Since the only textureless state is $f_1$, a pure state, the existence of a resource-destroying map is evident: It corresponds to the replacement map, taking all states to $f_1$. The map can be explicitly built and reads 
$$K_n=| f_1\rangle \langle f_n|.$$ 
It is clear that $\sum_nK_n^{\dagger}K_n=\mathds{1}$ and $\Delta(\varrho)=\sum_nK_n\varrho K_n^{\dagger}=f_1$ for arbitrary $\varrho$, which fulfill (i) and (ii), simultaneously. $\blacksquare$

The existence of a resource-destroying map is important because it immediately implies that QST is not only a convex resource theory but also an affine one, see \cite{gour}. Proposition 7 places QST, along with quantum coherence, athermality, and asymmetry, within the most stringent category of resource theories \cite{gour}.

We remark that although the map is unique, it has infinitely many realizations via distinct sets of Kraus operators. For any orthonormal basis $\{ |f_1\rangle, |u_2\rangle, \cdots , |u_D\rangle \}$, and Kraus operators $\tilde{K}_1=|f_1\rangle \langle f_1|$, $\tilde{K}_2=|f_1\rangle \langle u_2|$, $\cdots$, $\tilde{K}_D=|f_1\rangle \langle u_D|$, the resulting map is $\Delta$. Of course, the QST-destroying map is necessarily non-unital: it takes the identity operator to $f_1$. For a detailed description of destroying maps, see section III.D.3 of \cite{gour}.

\section{Experimental Details}

\begin{figure}[!h]
\centering
\includegraphics[width=0.85\columnwidth]{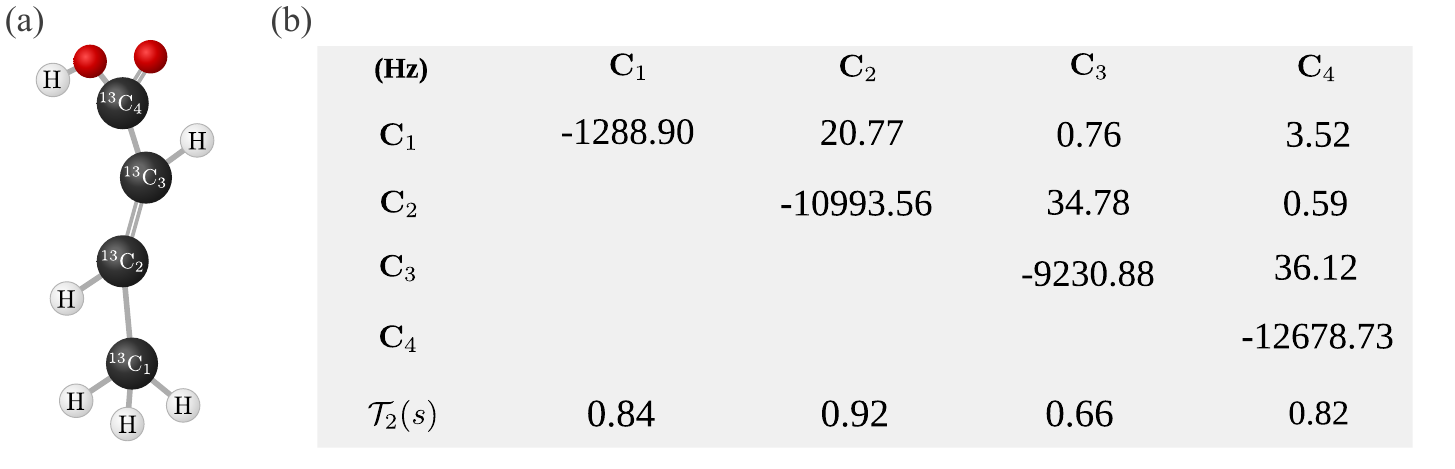}
\caption{(a) Molecular structure of $^{13}$C-labeled \textit{trans}-crotonic acid. (b) Hamiltonian parameters. The diagonal entries give the resonance offsets $\delta_{k}$, while the off-diagonal entries give the scalar couplings $J_{kj}$, all in Hz. The relaxation time $\mathcal{T}_2$ for each ${}^{13}$C nuclear spin is also indicated.}
\label{figSM01}
\end{figure}

\textit{NMR system}\textemdash Nuclear magnetic resonance (NMR) has proven to be an outstanding technique for proof-of-principle experiments in quantum information science~\cite{JONES202449,Ivan_book,Levitt2008,Mahesh_review,Vieira2023}. We conducted our experiments employing a liquid sample of $^{13}$C-labeled \textit{trans}-crotonic acid diluted in deuterated acetone. This molecule contains four coupled ${}^{13}$C nuclear spins, denoted as ${}^{13}\rm C_1$, ${}^{13}\rm C_2$, ${}^{13}\rm C_3$,${}^{13}\rm C_4$, which encode the system $S$ and ancillary $A$ qubits used throughout the experiments, see Fig.~\ref{figSM01}(a). The role assigned to each carbon spin depends on the implemented protocol, and we specify the corresponding system–ancilla assignments for each quantum circuit. We decouple all hydrogen atoms during all experiments. 

In the rotating frame and under the weak coupling approximation, the natural Hamiltonian of an $n$-qubit NMR system can be expressed as
\begin{equation}
\mathcal{H}_{0}=\sum_{k}\pi\delta_{k}\sigma_{z}^{k}+\sum_{k<j}\frac{\pi}{2}J_{kj}\sigma_{z}^{k}\sigma_{z}^{j},
\label{H0}
\end{equation}
where $\delta_{k}=\nu_k-\nu_{k}^{\text{rf}}$ is the frequency offset in Hertz, $\nu_k$ is the Larmor frequency of the $k$th nucleus, $\nu_{k}^{\text{rf}}$ is the rotating-frame frequency, $J_{kj}$ are the scalar $J$-coupling strength between the $k$th and $j$th nuclear spins, and $\sigma_z$ is the Pauli matrix~\cite{Ivan_book,Levitt2008,JONES202449}. In our setup, the nuclear spins precess around a strong static magnetic field of magnitude $B_0\approx 7.050$T, corresponding to a Larmor frequency of approximately $75.49$ MHz for the $^{13}$C nucleus (with a small offset of a few kHz due to the chemical shift). Figure~\ref{figSM01}(b) displays the measured offset magnitudes, the carbon scalar couplings, and the spin relaxation time extracted for this sample. The Larmor frequencies and coupling strengths are provided in the diagonal and off-diagonal positions, respectively.

In addition to the natural Hamiltonian ($\mathcal{H}_{0}$), we can apply radiofrequency pulses (RF) on resonance with the $k$th nuclear spin, resulting in the dynamics (in the rotating frame)
\begin{equation}
\mathcal{H}_{\rm RF}(t)=\frac{1}{2} \sum_{k}^{} \omega_{k}(t) \big[ 
\cos(\phi_k)\, \sigma_{x}^{k} 
+ \sin(\phi_k)\, \sigma_{y}^{k} \big],
\end{equation}
where \(\omega_k(t) = 2\pi\gamma_k B^{k}_1(t)\) is the time-dependent amplitude, with \(\gamma_k\) denoting the gyromagnetic ratio of the $k$th nucleus and $B_1(t)$ the time-dependent amplitude of the radiofrequency field. During NMR experiments, the phase \(\phi_k(t)\) and the amplitude $\omega^k(t)$ of the radiofrequency pulses are properly modulated to implement the desired quantum control operations on the nuclear spin qubits. 

\textit{Pseudopure state preparation}\textemdash The NMR system consists of a large ensemble of spins at room temperature~\cite{Ivan_book,Levitt2008}. In the high-temperature expansion, the system can be initially described by the thermal equilibrium state, 
\begin{equation}
\label{rho_eq}
\varrho_{eq}=\frac{\mathbb{I}}{16}+\varepsilon\Delta\varrho,
\end{equation}
where $\mathbb{I}$ is the $16 \times 16$ identity matrix,  $\varepsilon\approx10^{-6}$
is the thermal spin polarization, and $\Delta\varrho=(\varrho_{eq} - \mathbb{I}/16)/\varepsilon$ is the deviation matrix. Since the identity component is not experimentally accessible, all observed NMR signals arising from control-field manipulations are due solely to the traceless deviation matrix. Because the identity component does not contribute to the detected NMR signal, quantum-information experiments are conveniently described in terms of an effective pseudopure state. There are several different techniques used to initialize the NMR system, including the spatial averaging method~\cite{Cory_1997}, temporal averaging~\cite{Laflamme_PRA98}, among others~\cite{Ivan_book}. In our experiments, we employed the spatial-averaging approach, comprising single-qubit rotations, free evolution, and pulsed-field gradients, to initialize our system in the pseudopure state (PPS). In a PPS, the deviation density matrix is proportional to a pure-state projector,
\begin{equation}
\varrho_{pps}=\frac{(1-\varepsilon)}{16}\mathbb{I}+\varepsilon |0000\rangle\langle 0000|.
\label{PPS}
\end{equation}
Therefore, for the unitary manipulations and traceless observables considered here, the identity component is dynamically inert and experimentally invisible, so all the measured NMR signal can be described in terms of the effective pure deviation state $|0000\rangle\langle 0000|$.

\begin{figure}[!h]
\centering
\includegraphics[width=0.85\columnwidth]{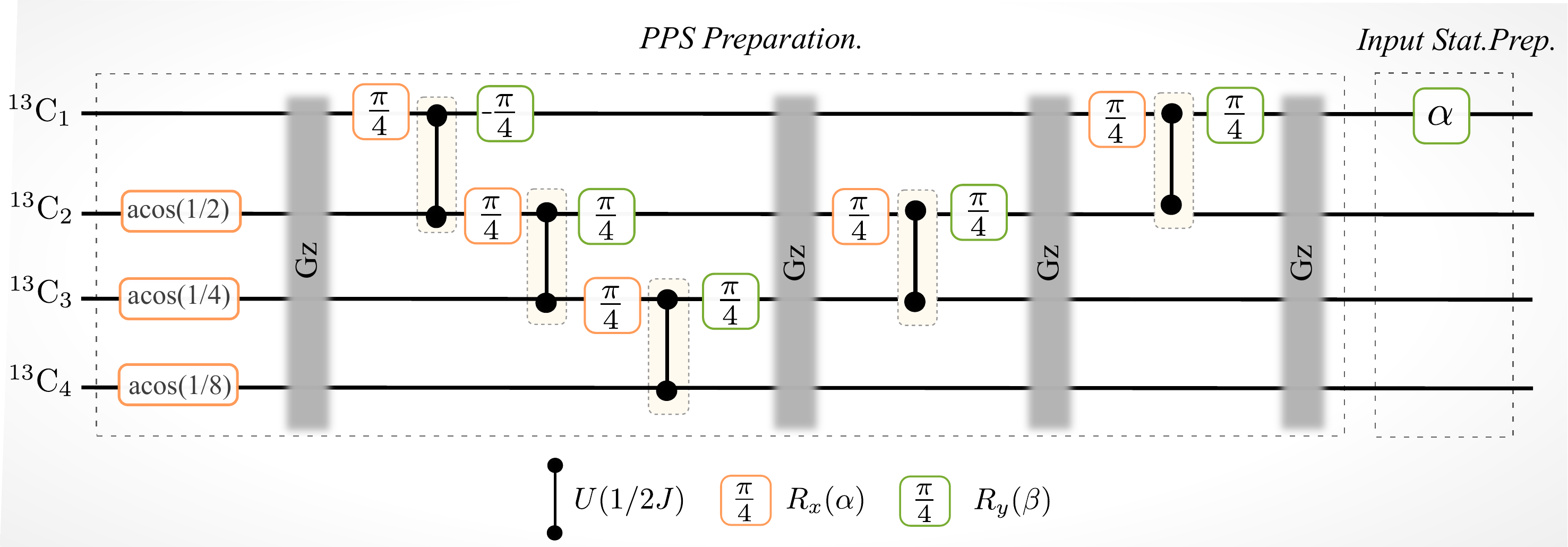}
\caption{Spatial-averaging NMR pulse sequence used to prepare the PPS on the register of four carbon spins. The orange and green rectangles indicate the $R_x$ and $R_y$ rotation gates, and $U(1/2J)$ denotes periods of free evolution, respectively. The gray rectangles denote the gradient-field pulses, $G_z$, applied to eliminate the coherence elements from the instantaneous deviation matrix.}
\label{figSM02}
\end{figure}

Figure~\ref{figSM02} illustrates the NMR pulse sequence implemented to prepare the $4$-qubit PPS state (Eq.~\ref{PPS}) starting from the thermal equilibrium state described by Eq.~(\ref{rho_eq}). The initial rotations redistribute the thermal populations, the controlled evolutions implement the required pairwise population transfers, and the pulsed field gradients, $G_z$, remove unwanted coherence elements. In our experiment, we combined each circuit segment, separated by four gradient-field pulses, into a single unitary evolution and used optimal control to design the corresponding RF pulses. We obtained shaped pulses with lengths of $3$ ms, $20$ ms, $15$ ms, and $15$ ms, respectively. All pulses were optimized to reduce the sensitivity to small inhomogeneities in the RF control fields and achieve high fidelity relative to the target pulses.

\textit{Input state preparation}\textemdash Following the PPS preparation discussed before, the joint system–ancilla register is effectively initialized in the state $\rho_{\mathrm{SA}}^{(0)}=|0\rangle\langle 0|_{\mathrm S}\otimes\left(|0\rangle\langle 0|_{\mathrm A}\right)^{\otimes 3}$. To generate the three input states employed throughout the experiments, we apply a single-qubit rotation \(R_y(\alpha)=e^{-i\alpha\sigma_y/2}\) to the $^{13}\text{C}_1$ nuclear spin. We set \(\alpha=0\), \(\pi/2\), \(-\pi/2\), preparing  \(|0\rangle\), \(|+\rangle=(|0\rangle+|1\rangle)/\sqrt{2}\), and \(|-\rangle=(|0\rangle-|1\rangle)/\sqrt{2}\), respectively, while the ancilla qubits remain in \(|000\rangle\). 

\begin{figure}[!h]
\centering
\includegraphics[width=0.95\columnwidth]{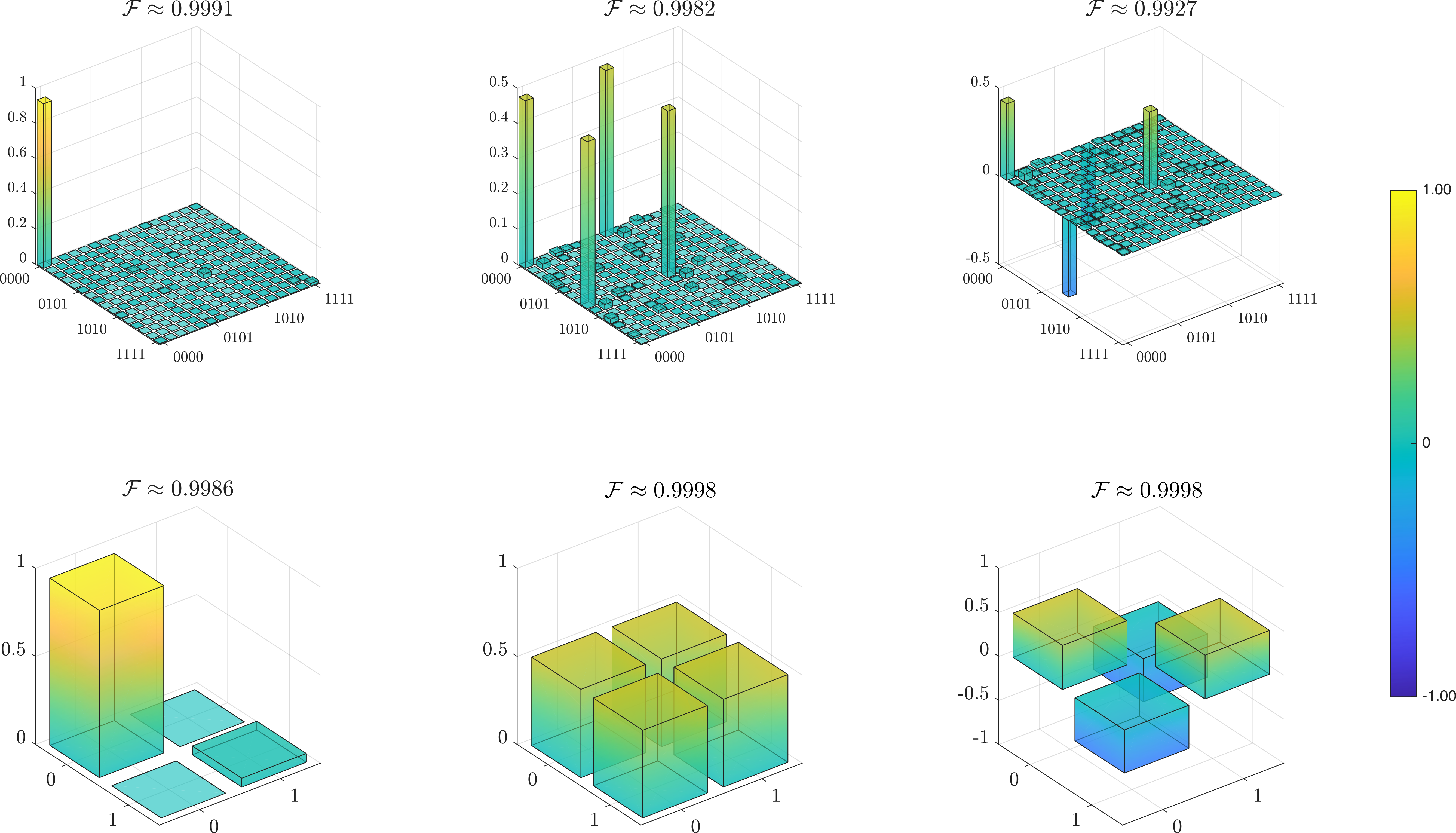}
\caption{Real parts of the experimentally reconstructed density matrices for the three input states used throughout the experiments. Upper row: full four-qubit states $\varrho_{\rm SA}$ of the system–ancilla register. Lower row: reduced states $\varrho_{\rm S}$ of the system qubit after tracing out the ancilla qubits. The states correspond to $|0\rangle$, $|+\rangle$, and $|-\rangle$ (left to right). The normalized Hilbert-Schmidt fidelities between experimental and theoretical states are approximately \(0.9991\), \(0.9982\), and \(0.9927\) for the full four-qubit register, and \(0.9986\), \(0.9998\), and \(0.9998\) for the reduced system qubit, respectively. The imaginary parts of all density matrices are negligibly small (on the order of $10^{-3}$), confirming the accuracy of the input state preparation. Small deviations from the ideal structure arise from experimental imperfections, including pulse errors and decoherence.}
\label{figSM03b}
\end{figure}

Figure \ref{figSM03b} presents the real parts of the experimentally reconstructed density matrices for these three input states. Full quantum state tomography of the four-qubit register was performed using $17$ tomographically complete measurement settings, each consisting of $\pi$ rotations along selected Pauli operator strings, from which the experimental density matrix $\varrho_{\rm exp}$ was reconstructed~\cite{Dawei_PRA20}. The upper row displays the density matrices of the complete system-ancilla register, while the lower row shows the corresponding reduced states, $\varrho_{\rm S}=\text{Tr}_{\rm A}(\varrho_{\rm SA})$, obtained by tracing out the three ancilla qubits. To quantify the quality of the state preparation, we compute the normalized Hilbert-Schmidt fidelity, $\mathcal{F}(\varrho_{\rm exp},\varrho_{\rm the})=\text{Tr}(\varrho_{\rm exp}\varrho_{\rm the})/\sqrt{\text{Tr}(\varrho_{\rm exp}^2)}\sqrt{\text{Tr}(\varrho_{\rm the}^2)}$ between the theoretical and experimental input states~\cite{Leskowitz_PRA04,LEE2002349}. For the full four-qubit register, the reconstructed states achieve fidelity of approximately \(0.9991\), \(0.9982\), and \(0.9927\) for the \(|0\rangle\), \(|+\rangle\), and \(|-\rangle\) inputs, respectively. After tracing out the ancilla qubits, the reduced system states attain fidelities of \(0.9986\), \(0.9998\), and \(0.9998\). The small residual matrix elements outside the expected populations and coherences arise from experimental imperfections and reconstruction uncertainty. In general, the observed density-matrix structure and high fidelities demonstrate accurate preparation of the three input states and confirm that the ancillary qubits remain effectively initialized in their reference state $|000\rangle$. 

\textit{Measurement}\textemdash In NMR, the measured signal results from an ensemble average over a large number of identical molecules. After an RF pulse rotates the magnetization into the transverse plane $(xy)$, the precessing spins induce an oscillating current in the probe coils at frequencies close to each nucleus's Larmor frequency. This process generates a non-equilibrium signal (Free Induction Decay, FID) that decays exponentially due to the relaxation processes. The FID corresponds to a time-domain signal and encodes the information regarding the transverse magnetization of each nuclear spin as,
\begin{equation}
\text{FID}(t) \varpropto \sum_{k}\text{Tr}[\varrho(t) \sigma^{(k)}_{-}]e^{-t/\mathcal{T}^{k}_{2}},\quad \sigma^{(k)}_{-}=(\sigma^{k}_x-i\sigma^{k}_y)/2,
\end{equation}
where $\varrho(t)=e^{-i\mathcal{H}_{0}t}\varrho_{0}e^{i\mathcal{H}_{0}t}$ is the evolved density matrix $\varrho_{0}$ under the natural NMR Hamiltonian, $\mathcal{H}_{0}$, and $\mathcal{T}^{k}_{2}$ is the natural spin relaxation time for each nuclear spin $k$ from which we are observing the signal. This signal includes both real and imaginary components, which correspond to the magnetization along the $x$ and $y$ axes and encode the expectation values of the Pauli matrices $\sigma_x$ and $\sigma_y$ for each observed spin $k$, respectively. During the experiments, we acquire data by repeatedly running the experiment, with each repetition tuned to the specific nuclear spin frequency to be detected.  Although only one part in a million of the molecules has the desired magnetization, the large number of molecules in the sample produces an excellent signal-to-noise ratio. Thus, each acquired FID signal represents a spatial average over many identical molecules.

\begin{figure}[!h]
\centering
\includegraphics[width=\columnwidth]{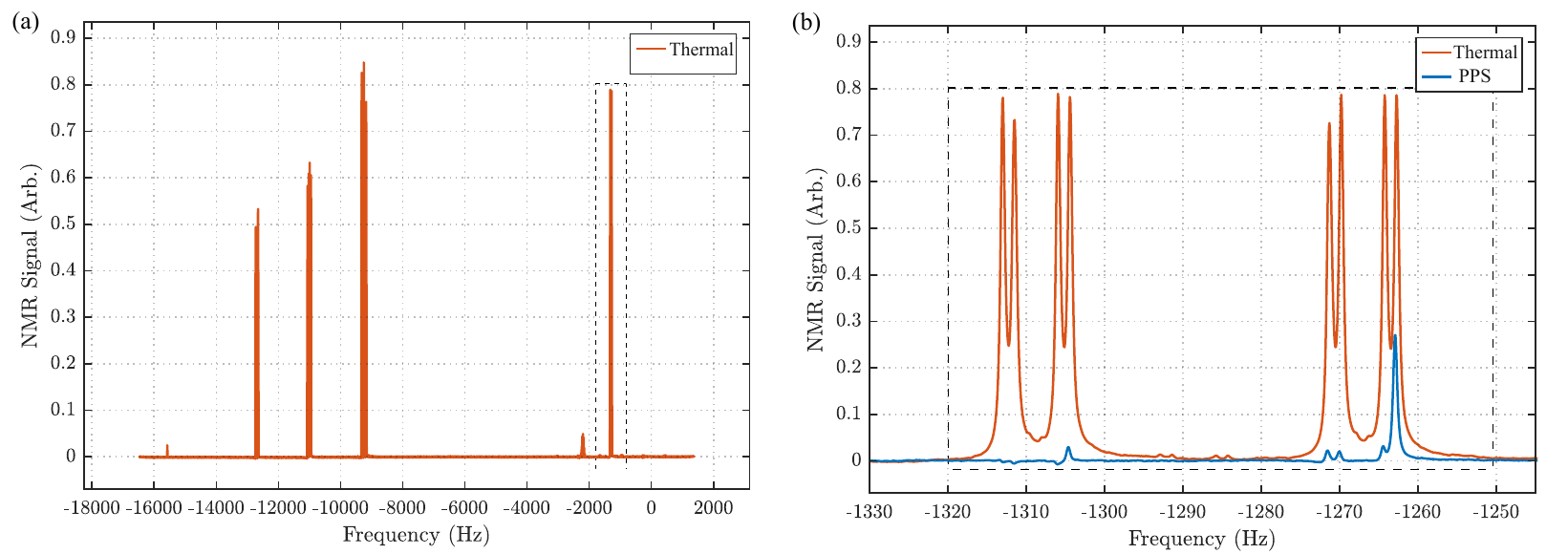}
\caption{(a) Experimental $^{13}\rm C$ NMR spectrum of crotonic acid at thermal equilibrium. (b) Zoom around the $^{13}\text{C}_1$ multiplet showing the PPS spectrum (blue line) superimposed on the thermal spectrum (red line), demonstrating the suppression of unwanted thermal lines after the PPS preparation. The PPS spectrum
was obtained by performing a $\pi/2$ readout pulse on the $^{13}\text{C}_1$ spin after the pulse sequence displayed in Fig~\ref{figSM02} and before data acquisition.}
\label{figSM03}
\end{figure}

For the \textit{trans}-crotonic acid molecule used as a four-qubit NMR quantum processor, the FID can be expressed as the sum of the transverse signals generated by the four $^{13}\mathrm{C}$ nuclear spins. Each scalar coupling $J_{kj}$ introduces an additional oscillatory component to the FID. Upon Fourier transformation, we obtain the NMR spectrum. In the weak-coupling spectrum, each of the four $^{13}\mathrm{C}$ resonances is split into an eight-line multiplet by the three remaining coupled spins, yielding up to 32 resolved transitions, as shown in Fig.~\ref{figSM03}(a). The scalar couplings $J_{kj}$ can be experimentally determined from the frequency spacing between these peaks. See the corresponding experimental values listed in Fig.~\ref{figSM01}(b). Figure~\ref{figSM03}(b) compares the $^{13}\rm C$ NMR spectrum of the PPS obtained via spatial averaging (blue line) with the thermal equilibrium spectrum (red line). The unwanted thermal lines are effectively suppressed, yielding a spectrum consistent with the PPS deviation density matrix, which serves as the effective pure initial state for all subsequent control protocols

In all experiments, the central measured quantity is the grand sum, $\Sigma(\varrho)=D\langle f_1|\varrho|f_1\rangle$, which for a spin-1/2 system can be obtained directly from the transverse magnetization $\Sigma(\varrho)=1+\langle\sigma_x\rangle$, without requiring full quantum-state tomography. To extract the $x$-component of the transverse magnetization for the system qubit, we first normalized the experimental spectrum to a pseudopure-state reference spectrum. The multiplet associated with the observed $^{13}\text{C}_1$ spin, consisting of eight transitions arising from scalar couplings to the three remaining carbon spins, was then fitted using resonance frequencies determined from the calibrated NMR Hamiltonian. Each transition was modeled with absorptive and dispersive line-shape components, with a common transverse relaxation parameter and a global frequency offset included in the fit. The fitted absorptive amplitudes were then combined according to the Pauli-string transition pattern to obtain the expectation value of the desired observable. In particular, for the single-qubit $x$-magnetization measurement, the relevant observable is $\sigma_x\mathds1\mathds1\mathds1$, for which all eight absorptive contributions enter with the same sign. Therefore, the resulting normalized spectral amplitude yields the Pauli expectation value $\langle\sigma_x\rangle$, from which the system grand sum is directly obtained.

\textit{Experimental uncertainty analysis}\textemdash The main experimental imperfections in the NMR implementation arise from residual fluctuations and inhomogeneities in the RF control fields, deviations from the intended pulse modulation, and weak spatial inhomogeneities of the static and gradient magnetic fields. We optimized the control pulses to reduce sensitivity to these effects and achieve high-fidelity coherent control. The error bars reported in the experimental figures quantify the statistical uncertainty associated with the NMR spectral analysis. They were obtained through a Monte Carlo procedure in which Gaussian fluctuations, with variance estimated from the residuals of the corresponding spectral fits, were sampled and propagated through the complete spectral fitting procedure. We took the standard deviation of the resulting distribution of each observable as its statistical uncertainty and propagated these uncertainties to derived quantities using standard error propagation.

\section{Experimental implementation and characterization of quantum channels}

\subsection{Phase-damping channel}
The phase-damping (PD) channel $\Gamma_{\rm PD}$ describes the loss of quantum coherence without energy exchange between the system and its environment~\cite{nielsen2012quantum,breuer2007theory,schlosshauer2007}. Figure~\ref{figSM04}(a) shows the quantum circuit used to implement the PD channel. We initialize the system and ancilla qubits as discussed before. The evolution characterizing the channel is implemented by the global controlled operation, $U_{y}(\theta)=|0\rangle\langle0|_{S}\otimes I_{A}+|1\rangle\langle1|_{S}\otimes R_{y}(\theta)$, where $\theta$ is the rotation angle applied to the ancilla qubit. After this evolution and tracing out the ancillary qubit, the reduced dynamics of the system becomes
\begin{equation}
\Gamma_{\text{PD}}(\varrho)=\frac{1+\cos(\theta/2)}{2}\varrho+\frac{1-\cos(\theta/2)}{2}\sigma_{z}\varrho\sigma_{z},
\end{equation}
where the coherence elements are attenuated by $\cos(\theta/2)$. In turn, the action of this PD channel can be effectively described in terms of the Kraus operators
\begin{equation}
E_{0}=\left(\begin{array}{cc}
1 & 0\\
0 & \sqrt{e^{-\lambda}}
\end{array}\right);\;E_{1}=\left(\begin{array}{cc}
0 & 0\\
0 & \sqrt{1-e^{-\lambda}}
\end{array}\right),
\end{equation}
introducing the PD strength parameter $\lambda=-2\ln[\cos(\theta/2)]$, such that $\cos(\theta/2)=\exp[-\lambda/2]$. This parametrization was adopted so that equally spaced values of the dephasing parameter $\lambda$ correspond to the experimental control angles implemented through the controlled $R_{y}(\theta)$ rotation. Throughout all the experiments, the channels were carried out using the control angles $\theta=\{0, 1.84, 2.39, 2.69, 2.87, 2.98, 3.04, 3.08\}\,\rm rad$, corresponding to $\lambda=\{0,1,2,3,4,5,6,7\}$. 
\begin{figure}[!h]
\centering
\includegraphics[width=0.90\linewidth]{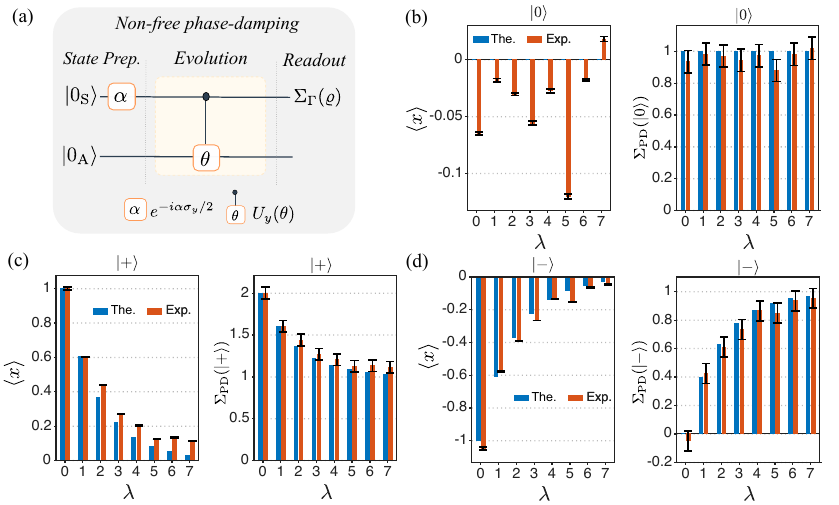}
\caption{Experimental measurement of the grand sum under a non-free phase-damping channel. (a) Schematic NMR circuit implementing the channel. (b-d) Measured transverse magnetization $\langle\sigma_x\rangle$ and corresponding grand sum $\Sigma_{\rm PD}(\varrho)=1+\langle\sigma_x\rangle$ for different input states $|0\rangle$, $|+\rangle$ and $|-\rangle$, respectively. Red bars denote experimental values, while blue ones are the theoretical predictions.}
\label{figS5b}
\end{figure}

According to the QST resource framework~(see Fig.1 in the main text), the PD channel is unital but not free. Under this map, the grand sum for the qubit system evolves as
\begin{equation}
\Sigma_{\mathrm{PD}}(\varrho)=1+e^{-\lambda/2}
\left[\Sigma_{\mathrm{in}}(\varrho)-1
\right],
\label{GSPD}
\end{equation}
where $\Sigma_{\mathrm{in}}(\varrho)\equiv\Sigma(\varrho^{\mathrm{in}}_{S})$ and
$\Sigma_{\mathrm{PD}}(\varrho)\equiv\Sigma(\varrho_{\Gamma}^{\text{out}})$ are the input and output grand sums, respectively. Although $\Gamma_{\text{PD}}$ is unital, it does not preserve the texture of individual states. Instead, regardless of the input state, the grand sum is driven toward the balanced value $\Sigma=1$ as shown in Fig.~\ref{figS5b}(b-d) for three states $|0\rangle$, $|+\rangle$ and $|-\rangle$, respectively.

For this channel, the dual $\tilde{\Gamma}_{\rm PD}$ transforms the textureless state according to
\begin{equation}
\tilde{\Gamma}_{\rm PD}(f_1) = \frac{1}{2}(\mathds{1}+e^{-\lambda/2}\sigma_x),
\end{equation}
and Proposition 1 therefore gives
\begin{equation}
\Sigma^{\prime}_{\mathrm{PD}}(\varrho)=1+e^{-\lambda/2}\langle\sigma_x\rangle_{\rm in},
\end{equation}
where $\langle\sigma_x\rangle_{\rm in}$ denotes the Pauli expectation value obtained from the input state $\varrho^{\text{in}}_{\text{S}}$. In Fig.~\ref{figSM04}(a), we show the residual difference $\Sigma_{\rm PD}-\Sigma^{\prime}_{\rm PD}$, which remains compatible with zero for all input states and dephasing strengths, providing a direct experimental consistency test of the dual-map description. Importantly, the two quantities compared in this experimental verification are obtained from independent data sets. The direct grand sum $\Sigma_{\rm PD}$ is extracted from the transverse magnetization measured after the implementation of the PD channel. At the same time, $\Sigma^{\prime}_{\rm PD}$ is evaluated from an independent measurement of the input-state magnetization.

\begin{figure}[!h]
\centering
\includegraphics[width=\linewidth]{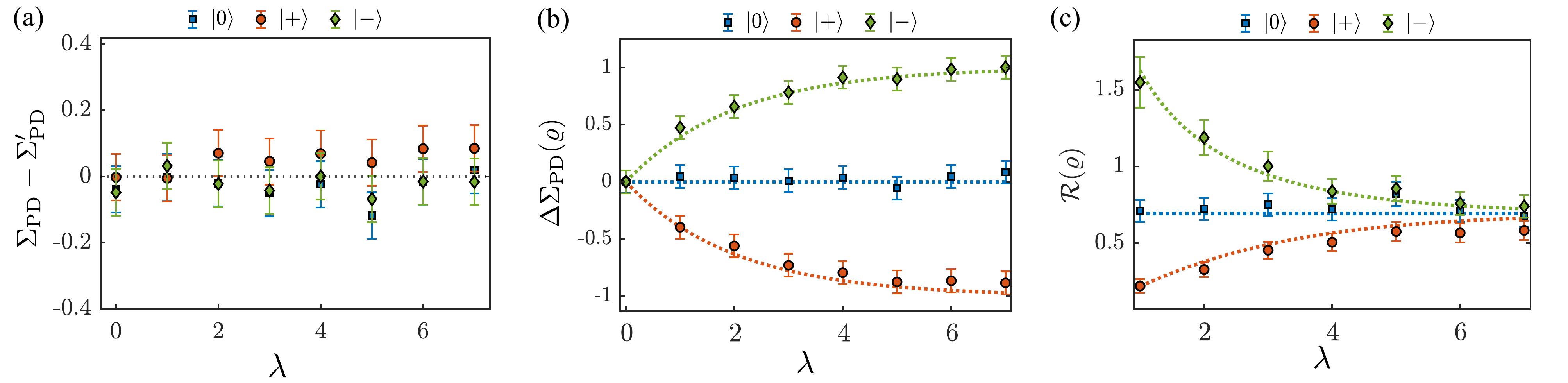}
\caption{QST dynamics under the non-free phase-damping channel. (a) Difference between the grand sum predicted from the dual evolution $\Sigma^{\prime}_{\rm PD}$ and the directly measured output grand sum $\Sigma_{\rm PD}$. The values remain compatible with zero within experimental uncertainty, providing a consistent test of Proposition 1 discussed in the main text. (b) Grand sum variation $\Delta\Sigma_{\rm PD}=\Sigma_{\mathrm{PD}}-\Sigma_{\mathrm{in}}$. States with $\Sigma_{\mathrm{in}}>1$ and $\Sigma_{\mathrm{in}}<1$ display opposite variations, while the state with $\Sigma_{\mathrm{in}}=1$ remains unchanged. (c) Corresponding rugosity, $\mathfrak{R}(\varrho)=-\ln[\Sigma(\varrho)/2]$, for the input states $|0\rangle$, $|+\rangle$, and $|-\rangle$, illustrating the convergence towards balanced value $\mathfrak{R}=\ln2$. The data are shown for $\lambda\geq1$, since in the ideal case state $|-\rangle$ has $\Sigma_{\rm PD}=0$ at $\lambda=0$, for which the rugosity diverges. Symbols are experimental data, and dotted curves are theoretical predictions.}
\label{figSM04}
\end{figure}

Figure~\ref{figSM04}(b) displays the grand sum variation
\begin{equation}
\Delta\Sigma_{\text{PD}}(\varrho)=\Sigma_{\mathrm{PD}}(\varrho)-\Sigma_{\mathrm{in}}(\varrho),  
\end{equation}
for the three input states under the non-free PD channel. The state $|+\rangle$, for which the ideal initial grand sum is maximal, $\Sigma_{\mathrm{in}}=2$, exhibits a negative variation, $\Delta\Sigma_{\rm PD}<0$, whereas $|-\rangle$, with $\Sigma_{\mathrm{in}}=0$, exhibits the opposite variation, $\Delta\Sigma_{\rm PD}>0$. The state $|0\rangle$, characterized by the balanced value $\Sigma_{\mathrm{in}}=1$, remains approximately unchanged, as expected from Eq.~\ref{GSPD}. Thus, the PD channel drives states on opposite sides of the balanced value toward $\Sigma=1$, explicitly revealing the redistribution of the grand sum induced by this unital but non-free dynamics.

Figure~\ref{figSM04}(c), shows the corresponding rugosity,
\begin{equation}
\mathfrak{R}(\varrho)=-\ln[\Sigma(\varrho)/2], 
\end{equation}
as a function of the dephasing strength. The initially textureless state $|+\rangle$, for which $\mathfrak{R}=0$ ideally, develops increasing rugosity as its grand sum decreases. Conversely, the initially highly textured state $|-\rangle$ exhibits a decreasing rugosity as its grand sum increases. Both evolve toward the balanced value $\Sigma_{\rm PD}=1$ corresponding to $\mathfrak{R}=\ln2$. The state $|0\rangle$, which already satisfies $\Sigma_{\rm PD}=1$, remains approximately unchanged throughout the evolution. These results show that unitality alone does not imply state-wise QST conservation: the non-free PD channel instead redistributes texture among different input states, consistent with the balanced redistribution underlying Proposition 4 of the main text.

\subsection{Free phase-damping channel}
\begin{figure}[!h]
\centering
\includegraphics[width=0.90\linewidth]{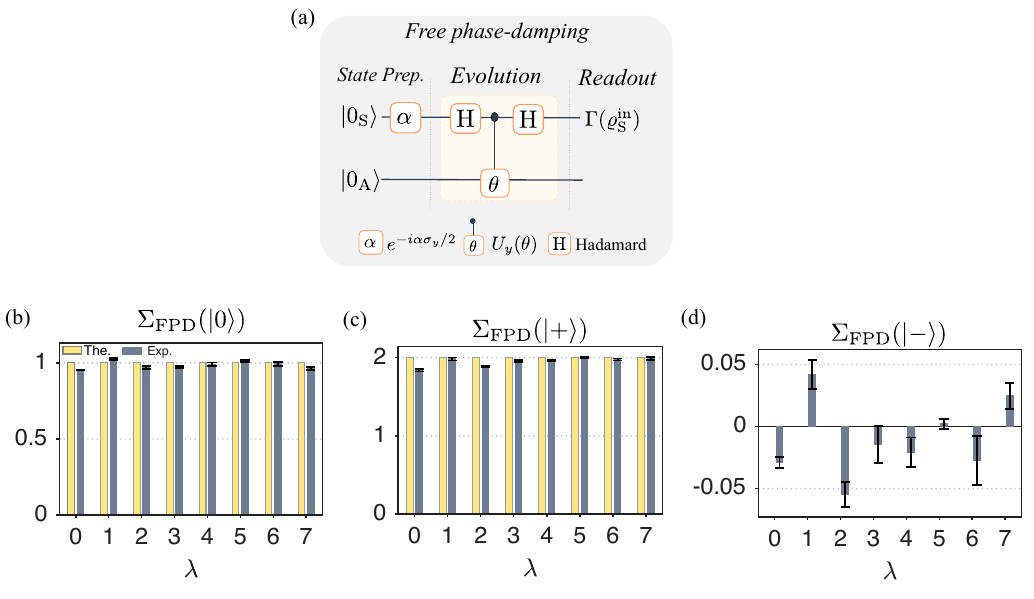} 
\caption{Experimental measurement of the grand sum under a free phase-damping channel. (a) Schematic NMR circuit implementing the FPD channel. (b-d) Experimental grand sum $\Sigma_{\rm FPD}(\varrho)$ dynamics for different input states $|0\rangle$, $|+\rangle$ and $|-\rangle$, respectively. Gray bars denote experimental values, while yellow ones are the theoretical predictions.}
\label{figSM07}
\end{figure}

We now address the free phase-damping (FPD) channel, whose quantum circuit is shown in Fig.~\ref{figSM07}(a). We obtain this channel by conjugating the phase-damping interaction with Hadamard gates acting on the system qubit. As a result, the effective dephasing occurs in the Hadamard basis $\{|+\rangle,|-\rangle\}$ rather than in the computational basis as discussed before. Since the textureless state for a qubit is $f_1=|+\rangle\langle+|$, this dynamics leaves $f_1$ invariant and therefore defines a free operation in the QST resource theory. After tracing out the ancillary qubits, the reduced dynamics of the spin system is given by,
\begin{equation}
\Gamma_{\text{FPD}}(\varrho)=\frac{1}{2}(1+e^{-\lambda/2})\varrho+\frac{1}{2}(1-e^{-\lambda/2})\sigma_{x}\varrho\sigma_{x},
\end{equation}
obeying $\Gamma_{\text{FPD}}(f_1)=f_1$ and ensures that $\Gamma_{\text{FPD}}$ is a free operation within the QST resource theory. Consequently, $\Sigma_{\rm FPD}(\varrho)=\Sigma_{\rm in}(\varrho)$, 
regardless of the input state. Figure~\ref{figSM07}(b-d) shows the evolution of the measured grand sum under the FPD channel for different initial input states. As expected from Proposition 3, the QST remains constant.

The dual map $\tilde{\Gamma}_{\rm FPD}$ acting in the textureless state can be written in the general single-qubit form
\begin{equation}
\tilde{\Gamma}_{\rm FPD}(f_{1})=\frac{1}{2}\left(m_{0}\mathds{1}+m_{x}\sigma_{x}+m_{y}\sigma_{y}+m_{z}\sigma_{z}\right),
\label{eq:dual_bloch_expansion}
\end{equation}
where the coefficients $m_\alpha$ define the Bloch representation of the operator $\tilde{\Gamma}_{\rm FPD}(f_{1})$. From the Proposition~2, the
channel preserves QST if and only if the textureless state is a fixed point of
the dual map, $\tilde{\Gamma}_{\rm FPD}(f_{1})=f_1$. Since $f_1=|+\rangle\langle+|$, the fixed-point condition is equivalent to
\begin{equation}
(m_0^{\rm th},m_x^{\rm th},m_y^{\rm th},m_z^{\rm th})
=(1,1,0,0).
\end{equation}

To experimentally test this condition, we reconstruct
$\tilde{\Gamma}_{\rm FPD}(f_1)$ from measurements of the grand sum. Based on Proposition~1, for any input state $\varrho_j$ we have
\begin{equation}
\Sigma_{\rm FPD}(\varrho_j)
=2\,{\rm Tr}
\left[\tilde{\Gamma}_{\rm FPD}(f_1)\varrho_j
\right].
\label{eq:prop1_fpd_reconstruction}
\end{equation}
By preparing the tomographically complete set of input states $\left\{
|0\rangle,\ |1\rangle,\ |+\rangle,\ |+i\rangle
\right\}$, with $|+i\rangle
\equiv(|0\rangle+i|1\rangle)/\sqrt{2}$, and measuring the corresponding output grand sum values
\begin{equation}
S_0\equiv\Sigma_{\rm FPD}(|0\rangle\langle0|),
\quad
S_1\equiv\Sigma_{\rm FPD}(|1\rangle\langle1|),
\end{equation}
\begin{equation}
S_+\equiv\Sigma_{\rm FPD}(|+\rangle\langle+|),
\quad
S_{+i}\equiv\Sigma_{\rm FPD}(|+i\rangle\langle+i|),
\end{equation}
we obtain a linear relation to experimentally reconstruct the coefficients of the operator $\tilde{\Gamma}_{\rm FPD}(f_{1})$
\begin{equation}
m_0=\frac{S_0+S_1}{2},
\qquad
m_z=\frac{S_0-S_1}{2},\qquad m_x=S_+-m_0,
\qquad
m_y=S_{+i}-m_0.
\label{eq:m_components_reconstruction}
\end{equation}

\begin{figure}[!ht]
\centering
\includegraphics[width=\linewidth]{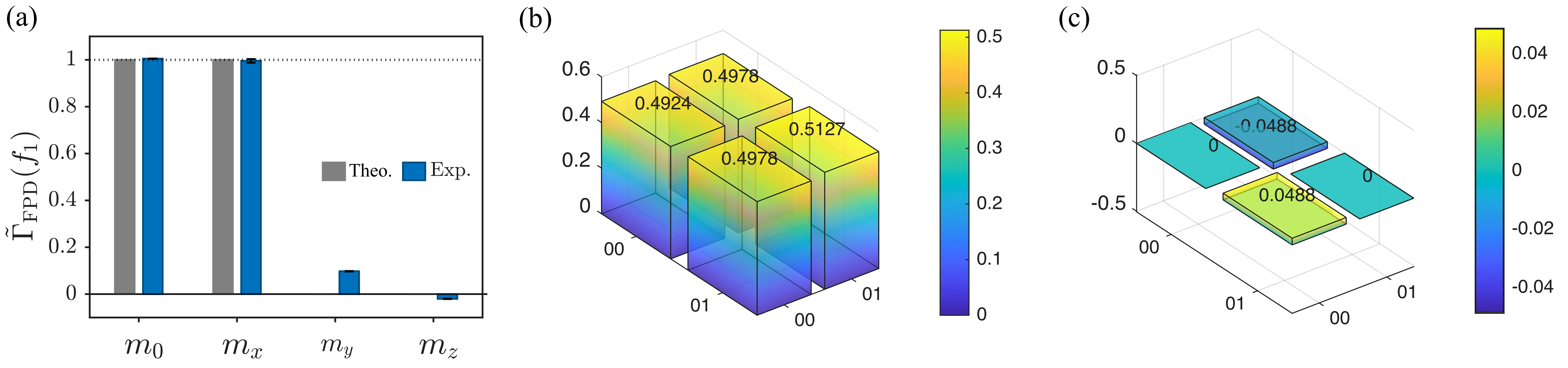}
\caption{Experimental verification of the dual fixed-point condition for the free phase-damping channel. (a) Reconstructed Pauli coefficients of
$\tilde{\Gamma}_{\rm FPD}(f_1)$, obtained from output grand sum measurements on a tomographically complete set of input states. Gray bars denote the theoretical fixed-point values $(1,1,0,0)$, and blue bars denote the experimental values. (b,c) Real and imaginary parts, respectively, of the reconstructed operator, showing agreement with the textureless state $f_1=|+\rangle\langle+|$. The results demonstrate that $\tilde{\Gamma}_{\rm FPD}(f_1)= f_1$, as required by Proposition~2 for QST-preserving dynamics.}
\label{figSM8a}
\end{figure}

Figure~\ref{figSM8a}(a) shows the experimentally reconstructed Pauli coefficients of the operator $\tilde{\Gamma}_{\rm FPD}(f_{1})$ for a fixed dephasing strength, $\lambda=4$. The blue bars denote the experimental
values, while the gray bars indicate the theoretical prediction $(1,1,0,0)$. The measured values are consistent with the theory within the experimental uncertainties: the identity and $\sigma_x$ components remain close to unity,
whereas the $\sigma_y$ and $\sigma_z$ components are compatible with zero. This 
experimentally validates the dual fixed-point condition of Proposition~2 and provides an operational signature of QST preservation. In Fig.~\ref{figSM8a}(b,c) we show the real and imaginary matrix representations of the reconstructed operator, respectively. As expected, the diagonal and off-diagonal elements of the real part are dominant and approximately equal to $1/2$, whereas the imaginary elements are small.

\begin{figure}[!ht]
\centering
\includegraphics[width=0.90\linewidth]{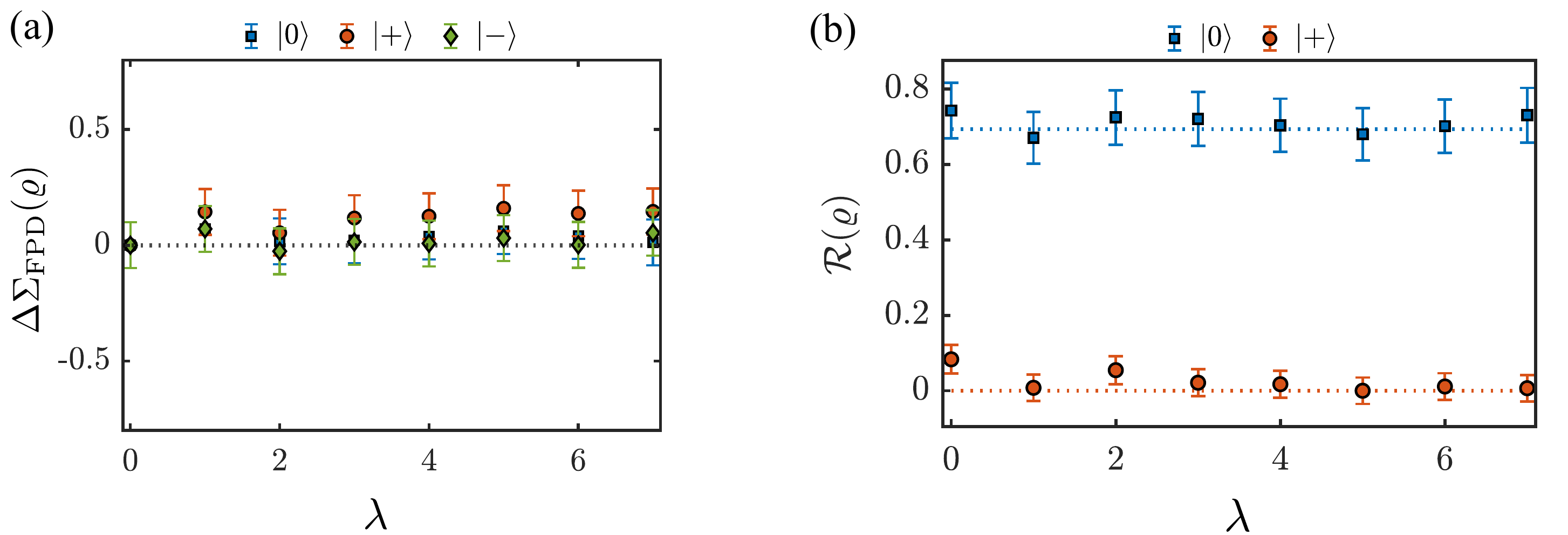}
\caption{QST conservation under the free phase-damping channel. (a) Grand sum variation $\Delta\Sigma_{\mathrm{FPD}}=\Sigma_{\mathrm{FPD}}-\Sigma_{\mathrm{in}}$. The values remain compatible with zero for all tested input states, as expected for a free unital operation. 
(b) Rugosity $\mathfrak{R}(\varrho)$ for $|0\rangle$ and $|+\rangle$, showing constant QST throughout the evolution in the FPD channel. The state $|-\rangle$ is omitted because $\Sigma_{\rm FPD}=0$ is preserved, making rugosity divergent in the ideal case. Symbols denote experimental data, and dotted lines represent theoretical predictions.}
\label{figSM05}
\end{figure}

Figure~\ref{figSM05}(b) illustrates the experimentally measured grand sum variation $\Delta\Sigma_{\text{FPD}}(\varrho)=\Sigma_{\mathrm{FPD}}(\varrho)-\Sigma_{\mathrm{in}}(\varrho)$ under the FPD channel. In contrast to the PD channel in Fig.~\ref{figSM04}(b), the FPD channel preserves the grand sum of each state throughout the evolution, i.e., $\Delta\Sigma_{\text{FPD}}(\varrho)=0$. The grand sum of all input states is preserved within experimental uncertainty, demonstrating no texture redistribution. This behavior experimentally verifies the prediction of Prop. 3 for the implemented free-unital channel. The same QST preservation can be viewed through the rugosity, $\mathfrak{R}(\varrho)$. As shown in Fig.~\ref{figSM05}(c), the states $|0\rangle$ and $|+\rangle$ keep approximately constant rugosity throughout the evolution, with ideal values $\mathfrak{R}=\ln2$ and $\mathfrak{R}=0$, respectively. The state $|-\rangle$ is not shown in the rugosity panel because the FPD channel preserves $\Sigma(|-\rangle)=0$, making the rugosity divergent in the ideal case.

\newpage
\subsection{Free amplitude-damping channel}
\begin{figure}[!h]
\centering
\includegraphics[width=0.90\linewidth]{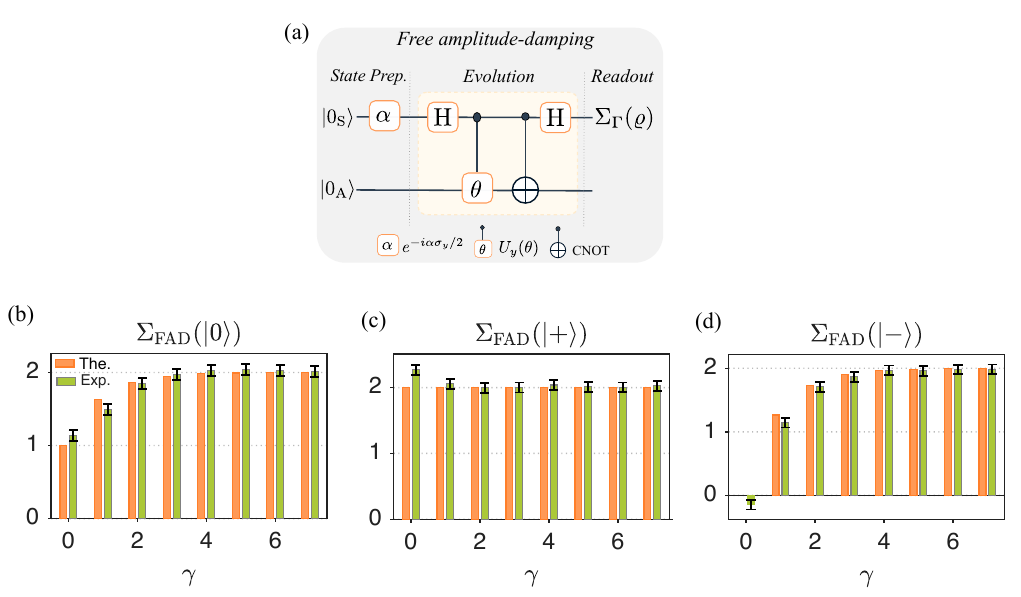}
\caption{Experimental measurement of the grand sum under a free amplitude-damping channel. (a) Schematic NMR circuit implementing the FAD channel. 
(b-d) Experimental dynamics of the measured grand sum under the FAD map for different input states. Green bars are experimental data, while orange ones denote the theoretical predictions.}
\label{figSM10}
\end{figure}

We next consider a free amplitude-damping (FAD) channel $\Gamma_{\rm FAD}$, a free but non-unital map in the QST resource theory. This process corresponds to amplitude damping in the
Hadamard basis of the system. Employing the quantum circuit illustrated in Fig.~\ref{figSM10}(a), the effective Kraus operators can be written as $L_{0}=|+\rangle\langle+|+\cos(\theta/2)|-\rangle\langle-|$ and $L_{1}=\sin(\theta/2)|+\rangle\langle-|$, and therefore 
the textureless state $f_1=|+\rangle\langle+|$ is a fixed point of the channel, $\Gamma_{\rm FAD}(f_1)=f_1$. The reduced dynamics of the system is given by
\begin{equation}
\Gamma_{\rm FAD}(\varrho)=\frac{1}{2}
\left(e^{-\gamma/2}+e^{-\gamma}\right)\varrho+\frac{1}{2}\left(e^{-\gamma}-e^{-\gamma/2}\right)
\sigma_x\varrho\sigma_x+\frac{1}{2}
\left(1-e^{-\gamma}\right)\left(\mathds{1}+\sigma_x\right),
\end{equation}
where $\gamma=-2\ln[\cos(\theta/2)]$ is the damping-strength parameter. In turn, the FAD maps the grand sum according to
\begin{equation}
\Sigma_{\rm FAD}(\varrho)=2-\left[2-\Sigma_{\rm in}(\varrho)\right]e^{-\gamma}.
\label{GS_FAD}
\end{equation}
Note that although $f_1$ is a fixed point of the channel, the FAD channel does not preserve the grand sum for arbitrary input states, reflecting its non-unital character. The dual map $\tilde{\Gamma}_{\rm FAD}$ takes the textureless state to
\begin{equation}
\tilde{\Gamma}_{\rm FAD}(f_1)=\frac{1}{2}[(2-e^{-\gamma})\mathds{1}+e^{-\gamma}\sigma_x].
\label{EqS25}
\end{equation}
For $\gamma>0$, Eq.(\ref{EqS25}) gives $\tilde{\Gamma}_{\rm FAD}(f_1)\neq f_1$. This is consistent with Proposition 3b, since a free non-unital channel cannot preserve QST for a qubit.

Figure~\ref{figSM10}(b-d) shows the behavior of the experimentally measured grand sum under the FAD channel dynamics. As expected, the output grand sum for all states converges to the common value, $\Sigma_{\rm FAD}(\varrho)\approx2$, in agreement with the theoretical prediction in Eq.~(\ref{GS_FAD}).

\begin{figure}
\centering
\includegraphics[width=\linewidth]{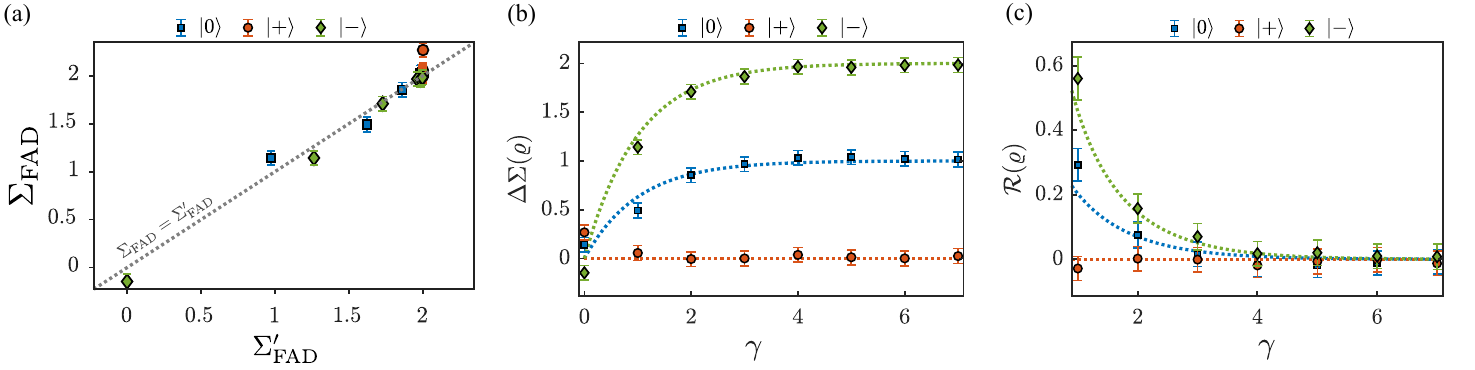}
\caption{Experimental characterization of the free amplitude-damping (FAD) channel. (a) Verification of Proposition~1 through the comparison between the directly measured grand sum, $\Sigma_{\rm FAD}$, and the dual-map prediction,
$\Sigma'_{\rm FAD}$. The gray dashed line indicates the identity $\Sigma_{\rm FAD}=\Sigma'_{\rm FAD}$. The accumulation of points near $\Sigma\simeq2$ reflects the relaxation toward the fixed state
$f_1=|+\rangle\langle+|$, whose grand sum is maximal.
(b) Grand sum variation
$\Delta\Sigma(\varrho)=\Sigma_{\rm FAD}(\varrho)-\Sigma_{\rm in}(\varrho)$
as a function of the damping strength $\gamma$. The $|+\rangle$ state remains
approximately invariant, whereas the $|0\rangle$ and $|-\rangle$ states show
positive shifts consistent with the FAD prediction.
(c) Rugosity $\mathcal{R}(\varrho)$ as a function of $\gamma$, showing the progressive smoothing of the state texture as the channel drives the system
toward $f_1$. Symbols denote experimental data and dotted curves represent the theoretical predictions.
}
\label{figSM11}
\end{figure}

Figure~\ref{figSM11} summarizes the experimental behavior of the free amplitude-damping (FAD) channel. Figure~\ref{figSM11}(a) tests Proposition~1 by comparing the directly measured grand sum, $\Sigma_{\rm FAD}$, with the independently constructed dual-map prediction, $\Sigma_{\rm FAD}'$. The data follow the identity line $\Sigma_{\rm FAD}=\Sigma_{\rm FAD}'$, confirming that the action of the channel on the grand sum is equivalently captured by the dual image of the textureless state. The clustering of several points near $\Sigma_{\rm FAD}\simeq\Sigma_{\rm FAD}'\simeq2$ is a characteristic feature of the FAD dynamics: as the damping strength $\gamma$ increases, the channel drives the system toward the fixed state $f_1$, whose grand sum is $\Sigma(f_1)=2$, making the output increasingly independent of the input state. Figure~\ref{figSM11}(b) shows the corresponding variation $\Delta\Sigma(\varrho)=\Sigma_{\rm FAD}(\varrho)-\Sigma_{\rm in}(\varrho)$. As expected from Eq.(\ref{GS_FAD}), the $|+\rangle$ state remains essentially unchanged, while the $|0\rangle$ and $|-\rangle$ input states exhibit positive grand sum variations. Finally, Figure~\ref{figSM11}(c) displays the rugosity $\mathcal{R}(\varrho)$, which decreases with $\gamma$, showing that the FAD channel progressively smooths the state texture as the dynamics thereby progressively depleting the QST~(QST-Depleting).

\subsection{Grand sum under composite free dynamics}We finally investigate QST dynamics under a continuously tunable family of free channels interpolating between a free non-unital (FNU) and a free-unital (FU) channel. Both elementary maps preserve the textureless state, $\Gamma(f_1)=f_1$, whereas only the FU component is unital. In the implementation used here, these channels correspond to FPD and FAD dynamics discussed before. The interpolation therefore enables us to continuously suppress the non-unital contribution while staying entirely within the set of free QST operations.

Considering the four-qubit circuit shown in Fig.4(a) of the main text, the control ancilla is initialized as
\begin{equation}
|\psi_{\beta}\rangle=\cos(\beta/2)|0\rangle_{\rm A}+\sin(\beta/2)|1\rangle_{\rm A}.
\end{equation} 
Its logical state selects the unitary dilation associated with either the FU or
the FNU process. The corresponding controlled evolution is
\begin{equation}
U_{\rm ctrl}(\theta_u,\theta_n)=|0\rangle\langle0|_{\rm A}\otimes U_{\rm FU}(\theta_u)+|1\rangle\langle1|_{\rm A}\otimes U_{\rm FNU}(\theta_n),
\end{equation}
where the elementary unitary blocks are
\begin{equation}
U_{\rm FU}(\theta_u)
=
\frac{1}{2}
\begin{pmatrix}
1+c_u & -s_u & 1-c_u & s_u\\
s_u & 1+c_u & -s_u & 1-c_u\\
1-c_u & s_u & 1+c_u & -s_u\\
-s_u & 1-c_u & s_u & 1+c_u
\end{pmatrix},
\; U_{\rm FNU}(\theta_n)
=
\frac{1}{2}
\begin{pmatrix}
1+c_n & -s_n & 1-c_n & s_n\\
s_n & 1+c_n & -s_n & 1-c_n\\
1-c_n & s_n & 1+c_n & -s_n\\
s_n & c_n-1 & -s_n & -(1+c_n)
\end{pmatrix},
\label{eq:UFU_block}
\end{equation}
with
\begin{equation}
c_u=\cos\theta_u,
\qquad
s_u=\sin\theta_u,
\qquad c_n=\cos\theta_n,
\qquad
s_n=\sin\theta_n. 
\end{equation}
After tracing out the control qubit and the ancillary degrees of freedom associated with the two unitary dilations, the reduced system dynamics is effectively described by the map
\begin{equation}
\Gamma_{\mu}(\varrho)
=
\mu\,\Gamma_{\rm FU}(\varrho)
+
(1-\mu)\,\Gamma_{\rm FNU}(\varrho),
\label{eq:composite_map}
\end{equation}
where $\mu=\cos^{2}(\beta/2)$ is controlled experimentally by the angle $\beta$ and tunes the relative contribution of the free-unital
and free non-unital branches. For $\mu=1$, the dynamics is purely free-unital, whereas for $\mu=0$ it reduces to the free non-unital channel. Since both elementary channels are free channels, the interpolating family also satisfies $\Gamma_{\mu}(f_1)=f_1$ for every $\mu$.

The resulting grand sum after
the composite free dynamics is 
\begin{equation}
\Sigma_{\Gamma_{\mu}}(\varrho)=\mu\,\Sigma_{\rm FU}(\varrho)+(1-\mu)\,\Sigma_{\rm FNU}(\varrho).
\end{equation} 
Since any free unital map preserves QST for arbitrary input states, i.e., $\Sigma_{\rm FU}(\varrho)=\Sigma_{\rm in}(\varrho)$, we have that 
\begin{equation}
\Delta\Sigma_{\mu}(\varrho)= (1-\mu)\Delta\Sigma_{\rm FNU}(\varrho).
\end{equation}
For the FNU amplitude-damping channel considered here $\Delta\Sigma_{\mu}(\varrho)$ generated by the composite free dynamics $\Gamma_{\mu}$ comes from its non-unital contribution. 

For the FNU amplitude-damping channel considered here,
\begin{equation}
\Delta\Sigma_{\mu}(\varrho)
=
(1-\mu)(1-e^{-\gamma})
\left[
2-\Sigma_{\rm in}(\varrho)
\right].
\end{equation}
In particular,
\begin{equation}
\Delta\Sigma_{\mu}(|+\rangle)=0,
\end{equation}
\begin{equation}
\Delta\Sigma_{\mu}(|0\rangle)
=
(1-\mu)(1-e^{-\gamma}),
\end{equation}
and
\begin{equation}
\Delta\Sigma_{\mu}(|-\rangle)
=
2(1-\mu)(1-e^{-\gamma}).
\end{equation}
The textureless state is invariant throughout the interpolation, whereas the
other input states progressively recover their initial grand sums as the
non-unital contribution is removed. In the limit $\mu\rightarrow1$,
\begin{equation}
\Delta\Sigma_{\mu}(\varrho)\rightarrow0
\end{equation}
for every input state, demonstrating the recovery of state-independent QST
conservation in the free-unital regime.

The experimental results in Fig. 4(b) of the main text follow these predictions. The measured grand sum variation $\Delta\Sigma_{\mu}(\varrho)$ decreases continuously as $\mu$ approaches one and becomes compatible with zero in the free-unital limit. The different amplitudes observed for the three input states reflect their different initial grand sum distances from the textureless fixed point. In particular, $|+\rangle$ is insensitive to the interpolation because it is a common fixed point of both elementary channels, whereas $|0\rangle$ and $|-\rangle$ display finite QST depletion whenever a non-unital component is present. These results provide a direct experimental illustration that, within the free-channel family implemented here, the non-unital FAD component drives the observed state-dependent variation of the grand sum.

\end{document}